\documentclass[conf]{new-aiaa}
\usepackage[utf8]{inputenc}

\usepackage{graphicx}
\usepackage{amsmath}
\usepackage[version=4]{mhchem}
\usepackage{siunitx}
\usepackage{longtable,tabularx}
\graphicspath{{images/}}
\usepackage{bm}
\usepackage{caption}
\usepackage{subcaption}
\usepackage{afterpage}
\usepackage{pdflscape}
\usepackage[table,xcdraw]{xcolor}
\usepackage{multirow}
\newcommand{\mymatrix}[1]{\left[\begin{array}{ccccccccccccccccccc} #1 \end{array}\right]}
\newcommand{\crossmat}[1]{\left[ #1 \times\right]}
\newcommand{\STAB}[1]{\begin{tabular}{@{}c@{}}#1\end{tabular}}
\newcommand{\coltit}[1]
{
    \begingroup
    \renewcommand{\arraystretch}{1} 
    \begin{tabular}[c]{@{}c@{}}#1\end{tabular}
    \endgroup
}
\newcommand{\newtab}{\end{tabular}\hspace{1cm}\begin{tabular}{ll}}

\title{Descent \& Landing Trajectory and Guidance Algorithms with Divert Capabilities for Moon Landing**}

\author{Francesco Capolupo\footnote{GNC Systems Engineer, Guidance, Navigation and Control Section, francesco.capolupo@esa.int.} and Antonio Rinalducci\footnote{GNC Systems Engineer, Guidance, Navigation and Control Section, antonio.rinalducci@esa.int. \newline **Author's original draft manuscript.}}
\affil{European Space Agency, ESTEC, 2201 AZ Noordwijk, The Netherlands}

\begin{document}

\maketitle

\begin{abstract}
This paper presents the preliminary design of the descent and landing trajectory and guidance algorithms of the ESA Argonaut lunar lander. The mission scenario and driving system constraints are presented and accounted for in the design of a fuel-optimal trajectory that includes divert capabilities, as required to achieve a safe landing. A sub-optimal descent and landing trajectory is then presented and computed from the optimal one, and the related on-board guidance algorithms are derived. The proposed end-to-end guidance solution represents an easily implementable alternative to on-board optimization, minimizing the verification \& validation effort, computational footprint, and programmatic risk in the development of the related GN\&C capabilities. A dedicated off-line optimization process is also outlined, and exploited to optimize the propellant consumption of the sub-optimal trajectory and to ensure the fulfilment of system constraints despite the use of simple algorithms on-board.
The sub-optimal trajectory is compared to the optimal baseline, and conclusions are drawn on the applicability of the proposed approach to the Argonaut mission.
\end{abstract}

\section*{Acronyms}
\noindent 
\begin{tabular}{ll}
     DE & Differential Evolution\\
     D\&L & Descent \& Landing \\              
     ELO & Elliptic Lunar Orbit \\              
     HDA & Hazard Detection and Avoidance  \\     
     LDE & Lunar Descent Element  \\             
     LGA & Low Gate \\
     MBB & Main Braking Burn
     \newtab
     MECO & Main Engine Cut-Off\\
     OCP & Optimal Control Problem\\
     PEG & Powered Explicit Guidance\\
     PGA & Pitch-up Gate\\
     RCS & Reaction Control Subsystem\\
     VGA & Vertical Gate\\
     V\&V & Verification \& Validation
 \end{tabular}

\section{Introduction}
\lettrine{T}{he} Argonaut Lander, formerly known as European Large Logistic Lander -- EL3, is a key element of the ESA Terrae Novae Exploration Programme. Its Lunar Descent Element (LDE) is planned to deliver scientific or logistic payload to the lunar surface starting from 2030. Its duties range from cargo delivery missions in support to the Artemis programme, to sample return and scientific/technology demonstration missions.
The objective of this work is to derive the optimal descent and landing trajectory, as well as to outline a possible on-board guidance solution and the related sub-optimal trajectory design process.

Argonaut is a $\sim$10 ton lander capable of delivering up to 1500 kg of payload with a safe pin-point landing. The foreseen GNC capabilities of the LDE include Terrain Relative Navigation (TRN), autonomous Hazard Detection and Avoidance (HDA), as well as throttleable engines. In the frame of this study, it is supposed that the lander is equipped with three 6 kN engines that can be throttled down to 50\% of their nominal thrust when needed, with a maximum thrust rate of 200 N/s per engine. In the nominal scenario, engines cannot be re-ignited once switched-off. A set of RCS thrusters is also included on the vehicle, to provide attitude control and limited position control capabilities. No thrust vectoring capabilities are foreseen for the main engines, instead RCS are used to steer the entire lander, and thus the engines' thrust, towards the required direction. A maximum steering rate of 5 deg/s is assumed for the computation of the descent and landing trajectory. 

\begin{figure}
    \centering
    \includegraphics[width=0.7\textwidth]{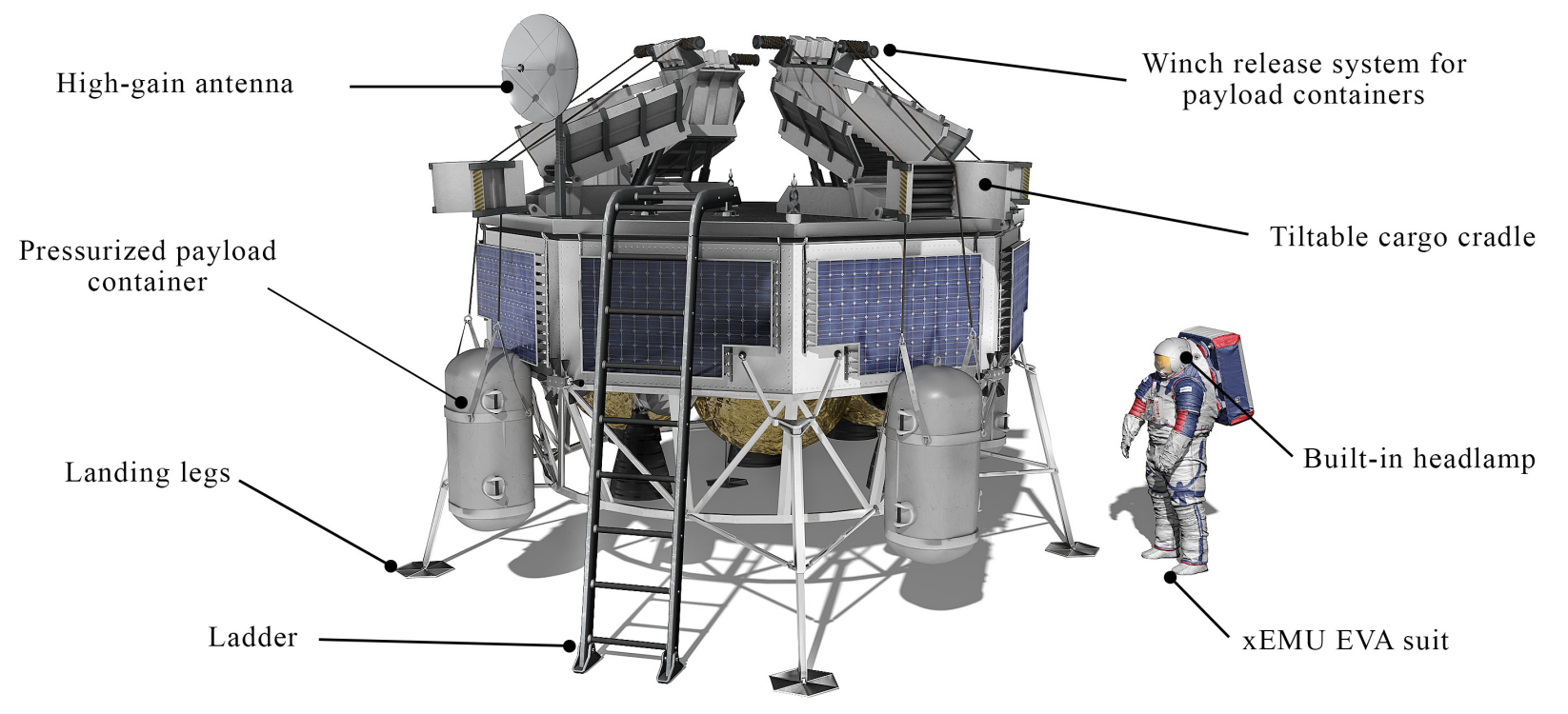}
    \caption{Hypothetical Argonaut lander configuration \cite{3dmodel}.}
    \label{fig:argoModel}
\end{figure}

\begin{figure}[b]
    \centering
    \includegraphics[width=0.8\textwidth]{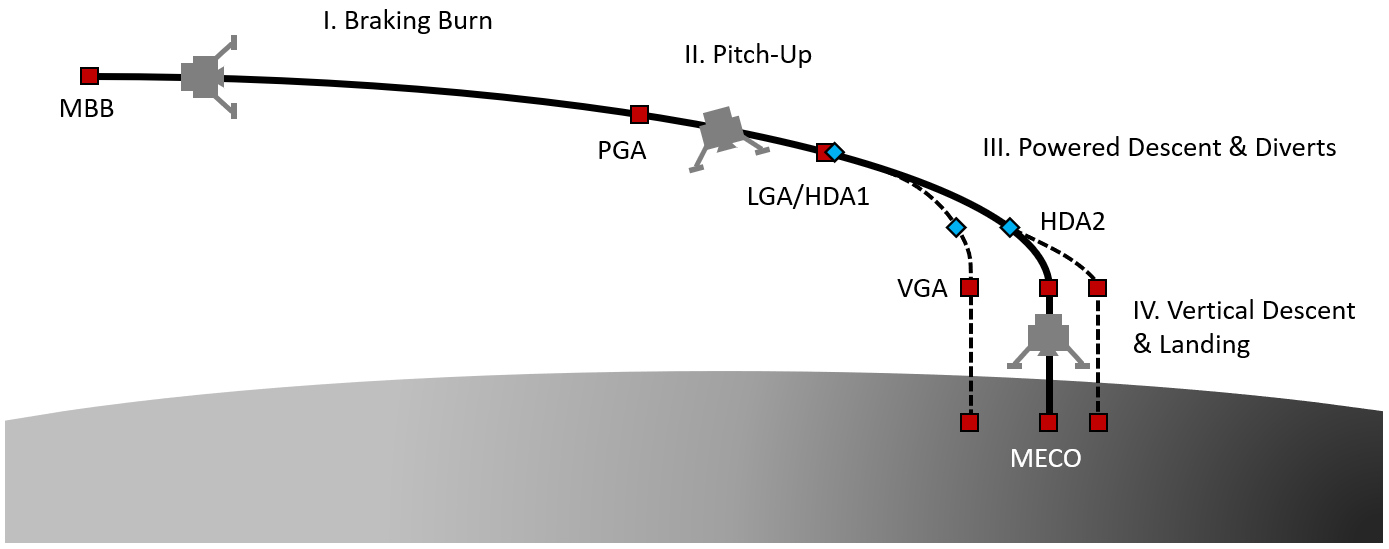}
    \caption{Descent and landing scenario.}
    \label{fig:dlScenario}
\end{figure}

The descent and landing phase starts 30 km above the lunar surface at the periselene of a 100$\times$30 km Elliptic Lunar Orbit (ELO) whose orbital plane contains the desired landing site at touchdown. The trajectory reproduces the common scheme usually followed for lunar landings since the Apollo era \cite{apollo}, comprising four main phases: a braking burn phase where the maximum thrust of main engines is used to significantly reduce the velocity of the vehicle; a pitch-up phase during which the thrust vector is reoriented to an almost vertical direction, allowing for the use of dedicated D\&L cameras and sensors; a powered descent for the last few hundreds of meters to reach a desired state above the landing site; and finally a vertical descent of few tens of meters at a given constant vertical speed to ensure a soft landing. Five waypoints, namely Main Braking Burn (MBB), Pitch-up Gate (PGA), Low Gate (LGA), Vertical Gate (VGA), and Main Engines Cut-off (MECO) delimit the beginning and the end of each phase. The current engine staging strategy foresees the use of three engines from MBB to LGA and two engines from LGA onward. The central engine is supposed to be instantaneously switched off at LGA. The proposed D\&L scenario and waypoints are shown in Figure \ref{fig:dlScenario} and summarized in Table \ref{tab:wp}.

\begin{table}[!h]
\begin{center}
\begin{tabular}{lcccccccc}
    \textbf{D\&L Phase} & \textbf{from} & \textbf{to} & \textbf{Nb. of engines}\\ \hline
    1. Braking burn &  MBB & PGA & 3 \\
    2. Pitch up & PGA & LGA & 3 \\ 
    3. Powered descent and diverts & LGA & VGA & 2\\
    4. Vertical descent and landing & VGA & MECO & 2\\\hline 
\end{tabular}
\end{center}
    \caption{Descent \& Landing phases and waypoints.}
    \label{tab:wp}
\end{table}

\begin{table}[!h]
\begin{center}
\begin{tabular}{lrl}
\textbf{Parameter} & \multicolumn{2}{c}{\textbf{Value}}\\ \hline
ELO periselene altitude & 30 & km  \\
ELO aposelene altitude & 100&  km  \\
Wet mass at MBB & 7000 & kg\\
Max engine thrust & 6000 & N\\
Min engine thrust & 3000 & N\\
Max engine throttle rate & 200 & N/s\\
Engine specific impulse & 330 & s\\
Number of engines &  3\\
Max steering rate & 5 & deg/s\\
\hline \end{tabular}\hspace{1cm}\begin{tabular}{lrl}
\textbf{Parameter} & \multicolumn{2}{c}{\textbf{Value}}\\ \hline
LGA/HDA1 altitude & 500 &  m\\
LGA/HDA1 min pitch angle & 80 & deg\\
LGA/HDA1 max velocity & 30 & m/s\\
LGA/HDA1 max lateral divert & 100 & m\\
HDA2 altitude & 150 & m\\
HDA2 max lateral divert & 20 & m\\
VGA altitude & 30 & m\\
Vertical descent pitch angle & 90 & deg\\
Vertical descent velocity & 2 & m/s\\ \hline
\end{tabular}
\end{center}
    \caption{Descent and landing parameters and constraints.}
    \label{tab:parconstr}
\end{table}

Argonaut trajectory can include up to two divert manoeuvres that can be autonomously triggered during the powered descent phase. The first one is located at LGA, at an altitude of 500 m, and the second one is located shortly before VGA, at an altitude of 150 m. The combined divert manoeuvres can shift the horizontal position of VGA to up to $\pm$ 120 m. The maximum lateral divert at LGA/HDA1 is equal to 100 m, while the maximum lateral divert at HDA2 is limited to 20 m. Some additional constraints are also enforced at LGA, to allow for the correct operation of HDA sensors: these include a maximum velocity of 30 m/s and a minimum pitch angle of 80 deg. 

All the above-mentioned hypotheses and mission parameters are summarized in Table \ref{tab:parconstr}, and constitute the basis for the definition of a feasible descent and landing trajectory. Unsurprisingly, all the trajectory design work presented in Section \ref{sec2} is focused on the computation of a constrained fuel-optimal trajectory minimizing the overall propellant consumption, as this directly affects to one of the key mission-level parameters of Argonaut: the deliverable payload mass. Sections \ref{sec3} and \ref{sec4} tackle the challenging question of how to concretely realize the trajectory in flight, and derive the simplest possible guidance solution without making the landing problem simpler than what really is. Finally, the two resulting trajectories, i.e., the fuel-optimal one and the sub-optimal one obtained with the proposed guidance approach, are compared in Section \ref{cmp}, to conclude on the applicability of the latter to Argonaut.

\section{Optimal trajectory design}\label{sec2}
\subsection{Problem formulation}
The design of a fuel-optimal trajectory can be formulated as a multi-phase Optimal Control Problem (OCP) and solved numerically either via indirect or direct trajectory optimization techniques \cite{betts}. 

\begin{figure}[b]
    \centering
    \includegraphics[width=0.40\textwidth]{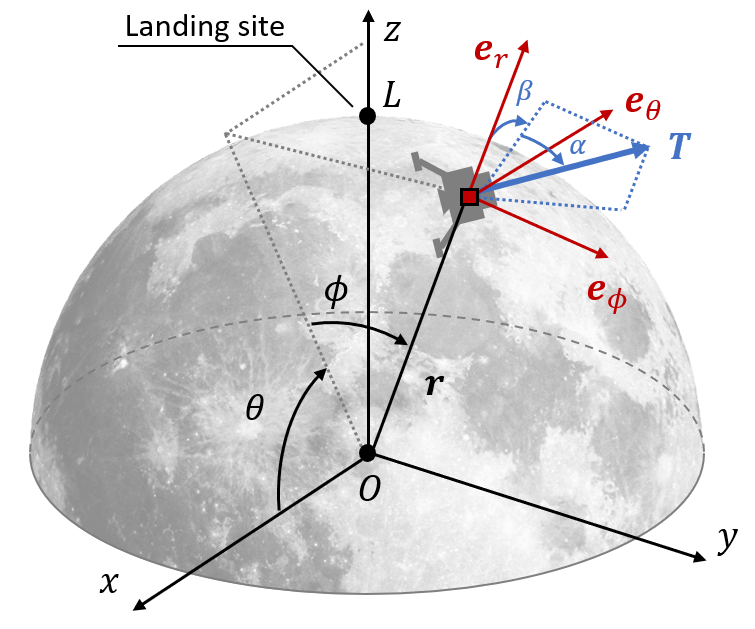}
    \caption{Spherical coordinates definition.}
    \label{fig:ldFrames}
\end{figure}

The translational dynamics of landing are modelled by a point mass in a Keplerian gravity field. We use a set of landing-site referenced spherical coordinates $\left(r,\phi,\theta\right)$, as defined in Fig. \ref{fig:ldFrames}, to describe the absolute position of the lander with respect to the center of the Moon
\[{\bm{r}} = r\left[\cos \phi \cos \theta \hspace{0.2cm}  \sin \phi \hspace{0.2cm}  \cos \phi \sin \theta\right]^\top\]
The thrust vector direction is parameterized by its azimuth $\alpha$ and elevation $\beta$ angles with respect to the $\left({\bm{e}}_r, {\bm{e}}_{\phi},{\bm{e}}_{\theta}\right)$ reference frame. The total thrust level is represented as the sum of the thrust $T_1$ provided by two outer engines, which stay on for the entire trajectory and must have the exact same thrust profile not to generate perturbing torques on the vehicle, and the thrust $T_2$ of the central engine, only used from MBB to LGA, and whose thrust profile is independent from the other two engines. We also suppose that all engines share the same constant specific impulse $I_s$. Under these hypotheses, the equations of motion are
\begin{subequations}
\begin{align}
\dot r & = v_r\\
\dot \phi & = \frac{v_{\phi}}{r}\\
\dot \theta & = \frac{v_{\theta}}{r \cos \phi}\\
\dot v_r & = \frac{v_{\phi}^2 + v_{\theta}^2}{r} - \frac{\mu}{r^2} + \frac{T_1+T_2}{m}\cos\alpha \cos\beta\\
\dot v_{\phi} & = -\frac{v_{\phi}v_r}{r} - \frac{v^2_{\theta}}{r}\tan\phi  + \frac{T_1+T_2}{m}\sin\alpha \\
\dot v_{\theta} & = -\frac{v_{\theta}v_r}{r} + \frac{v_{\phi}v_{\theta}}{r}\tan\phi + \frac{T_1+T_2}{m} \cos\alpha \sin \beta\\
\dot m & = -\frac{T_1 + T_2}{I_sg_0}
\end{align}
\label{polardyn}
\end{subequations}
where $v_r$, $v_{\phi}$, and $v_{\theta}$ are the projections of the inertial velocity vector in the $\left({\bm{e}}_r,{\bm{e}}_{\phi},{\bm{e}}_{\theta}\right)$ reference frame. 

The same exact phases of the D\&L trajectory are used as phases of the OCP, except for the vertical descent and landing phase, which is not included in the optimization problem as its trajectory is already completely specified by mission requirements. The objective is to minimize the propellant consumption of the main engines from MBB to VGA, or, equivalently, to maximize the vehicle mass at the end of the third phase (i.e., at the end of the powered descent phase)
\begin{equation}
    J = m_f^{\left(3\right)}
\end{equation}
It is important to underline that this formulation neglects the RCS propellant needed to steer the vehicle. No constraint is set on the argument of periselene of the ELO, nor on the final time at VGA, both considered as outcomes of the optimization process. Consequently, the boundary conditions of the problem are
\begin{subequations}
\begin{align}
    t_0^{\left(1\right)} & = 0 & t_f^{\left(3\right)} & = \text{free} \\
    r_0^{\left(1\right)} & = R_\text{Moon} + 30 \text{ km} & r_f^{\left(3\right)} & = R_\text{Moon} + 30 \text{ m}\\
    \phi_0^{\left(1\right)} & = 0 & \phi_f^{\left(3\right)} & = 0\\
    \theta_0^{\left(1\right)} & =  \text{free} & \theta_f^{\left(3\right)} & = 90 \text{ deg}\\
    v_{r_0}^{\left(1\right)}  & = 0 & v_{r_f}^{\left(3\right)}  & = -2 \text{ m/s}\\
    v_{\phi_0}^{\left(1\right)}  & = 0 & v_{\phi_f}^{\left(3\right)}  & = 0\\
    v_{\theta_0}^{\left(1\right)}  & = 1681.6 \text{ m/s} & v_{\theta_f}^{\left(3\right)}  & = 0 \\
    m_0^{\left(1\right)} & = 7000 \text{ kg} & m_f^{\left(3\right)} & = \text{free}
\end{align}
\end{subequations}

With this formulation we are implicitly neglecting the motion of the landing site due to the slow rotation of the Moon. In doing so we are potentially introducing an error of up to 5 m/s on the estimated $\Delta V$. This error is considered negligible in view of the approx. 1900 m/s needed to land. We ensure continuity of time, states, and commands across phases by introducing the following link constraints
\begin{subequations}
\begin{align}
t_f^{\left(i\right)} & = t_0^{\left(i+1\right)}\\
{\bm{x}}_f^{\left(i\right)} & = {\bm{x}}_0^{\left(i+1\right)}\\
\alpha_f^{\left(i\right)} & = \alpha_0^{\left(i+1\right)}\\
\beta_f^{\left(i\right)} & = \beta_0^{\left(i+1\right)}\\
T_{1,f}^{\left(i\right)} & = T_{1,0}^{\left(i+1\right)}
\end{align}
\end{subequations}
for $i = 1,2$. The central engine's thrust has a discontinuous profile at LGA, thus the only link constraint considered for thrust magnitude is the one at PGA
\begin{equation}
    T_{2,f}^{\left(1\right)} = T_{2,0}^{\left(2\right)}
\end{equation}
All the other constraints, notably those at LGA and VGA, are straightforwardly derived from Table \ref{tab:parconstr}. An additional constraint is introduced at VGA to ensure a smooth transition between powered and vertical descent
\begin{equation}
    \frac{T_{1,f}^{\left(3\right)}}{m_f^{\left(3\right)}} = \frac{\mu}{(r_f^{\left(3\right)})^2}
\end{equation}
requiring the outer main engines to equate the vehicle weight. Notice that there is no explicit constraint or condition imposed on PGA, which, from an optimization perspective, is a superfluous waypoint. We keep it anyways in our formulation, as it will become useful later when consolidating the nominal trajectory. It is also highlighted that, in the absence of out-of-plane divert manoeuvres, the trajectory optimization problem is purely planar (i.e., $\phi = v_{\phi} = \alpha = 0$).

\subsection{Nominal trajectory}
\begin{figure}
    \centering
    \includegraphics[width=\textwidth]{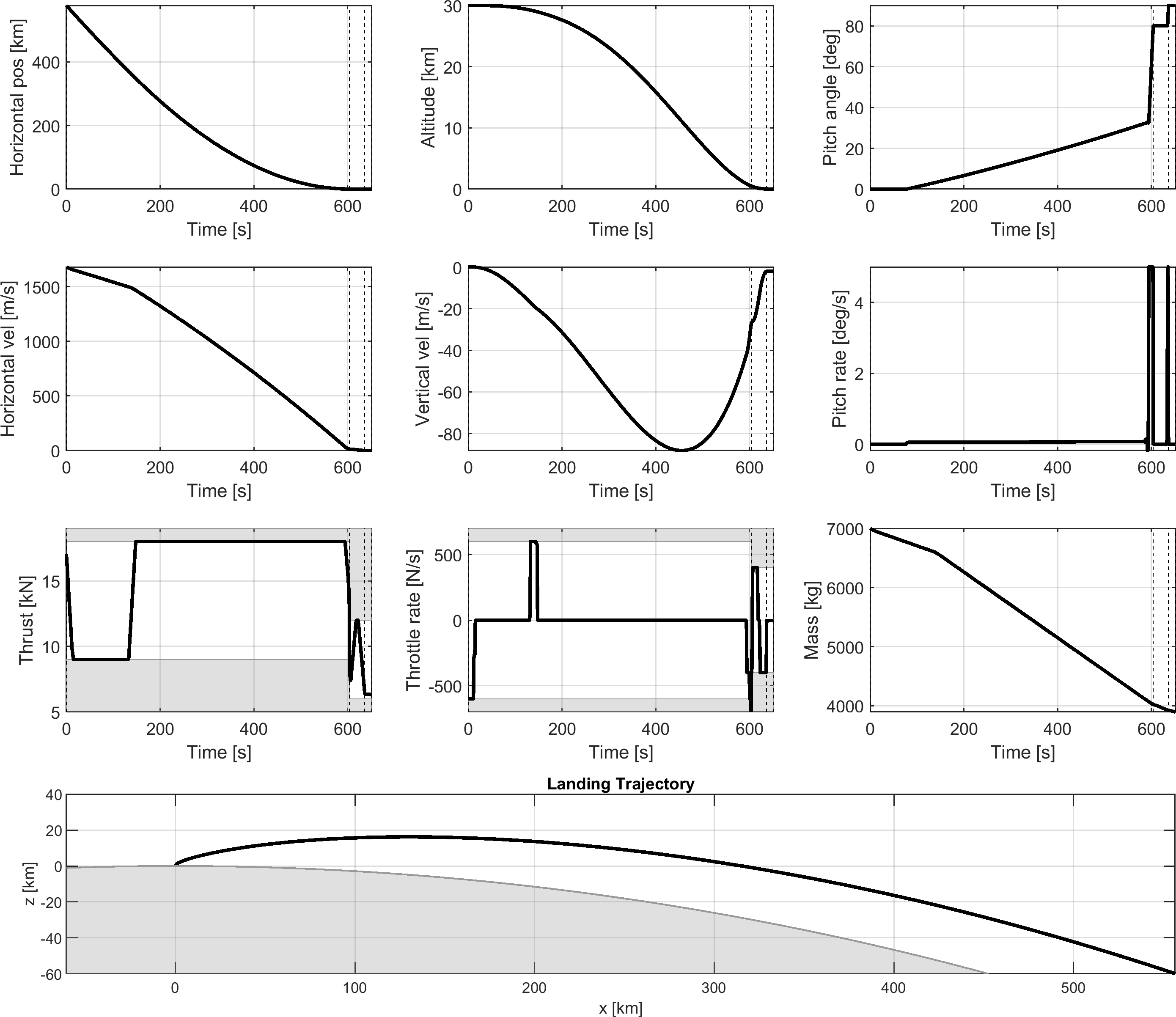}
    \caption{Optimal trajectory -- variable thrust during braking burn. Dotted vertical lines indicate the location of LGA, and VGA waypoints.}
    \label{fig:optTrajVarThrust}
\end{figure}

Figure \ref{fig:optTrajVarThrust} shows the solution obtained for the nominal optimal trajectory (i.e., no diverts) when solving the OCP by direct collocation techniques with IPOPT \cite{ipopt} as NLP solver. All key parameters of the trajectory, including states, pitch angle and rate, and thrust profile, are reported, together with the location of waypoints, which are indicated as dotted vertical lines. The bottom plot shows the entire descent and landing trajectory in an inertial reference frame having as origin the desired landing site. The optimal trajectory naturally follows the classic D\&L scheme outlined in the previous section: a first phase where all engines are fired for most of the time at their maximum thrust with a slowly drifting linear pitch angle profile, followed by a quick reorientation of the thrust direction and a significant reduction to the overall thrust level when reaching LGA, and then a slower powered descent to VGA which is concluded by another quick slew to acquire a 90 degree pitch angle. Notice that only two waypoints are visible on the plots: LGA at approximately 600 s and VGA right before the end of the trajectory. It turns out that, as previously anticipated, the optimizer does not exploit the PGA waypoint. Indeed, in the optimal trajectory PGA numerically coincides with LGA, even if, from a functional perspective, we could still identify PGA as the beginning of the first pitch manoeuvre at 5 deg/s. The resulting propellant mass consumption is equal to 3105.5 kg, corresponding to a $\Delta V$ of 1898.2 m/s, for a total time of flight of 651.5 s.

It is interesting to note that the thrust level profile starts close to its maximum value, then it is followed by a short idle phase at minimum thrust, and then increases again to its maximum value for the rest of the braking burn. This leads us to consider a second possible scenario where the thrust level is kept constant throughout the whole braking burn phase. Results for this alternative scenario are reported in Figure \ref{fig:optTraj}, showing that, in this case, the optimal braking burn strategy is to exploit all the available thrust. All the other parameters, including the thrust steering profile, are mostly unchanged with respect to the first optimal trajectory. This time we clearly see all internal waypoints: PGA, LGA, and VGA. PGA is actively exploited by the optimizer to reduce thrust and prepare for engine switch-off at LGA. The alternative trajectory yields a practically unchanged propellant consumption of 3105.8 kg, but a shorter time of flight of 584.4 s. It is also worth underlining that in this work we suppose that the specific impulse $I_s$ of the main engines does not depend on the thrust level. This is a simplifying hypothesis likely to be invalidated in real life. It is expected that $I_s$ decreases as the thrust level decreases, and, therefore, we can consider the mass consumption obtained for an idle thrust level as slightly underestimated. Consequently, and in view of the shorter time of flight, the constant thrust strategy is proposed as baseline for Argonaut. Its detailed timeline is reported in Table \ref{tab:optwp}.

\afterpage{
\begin{figure}
    \centering
    \includegraphics[width=\textwidth]{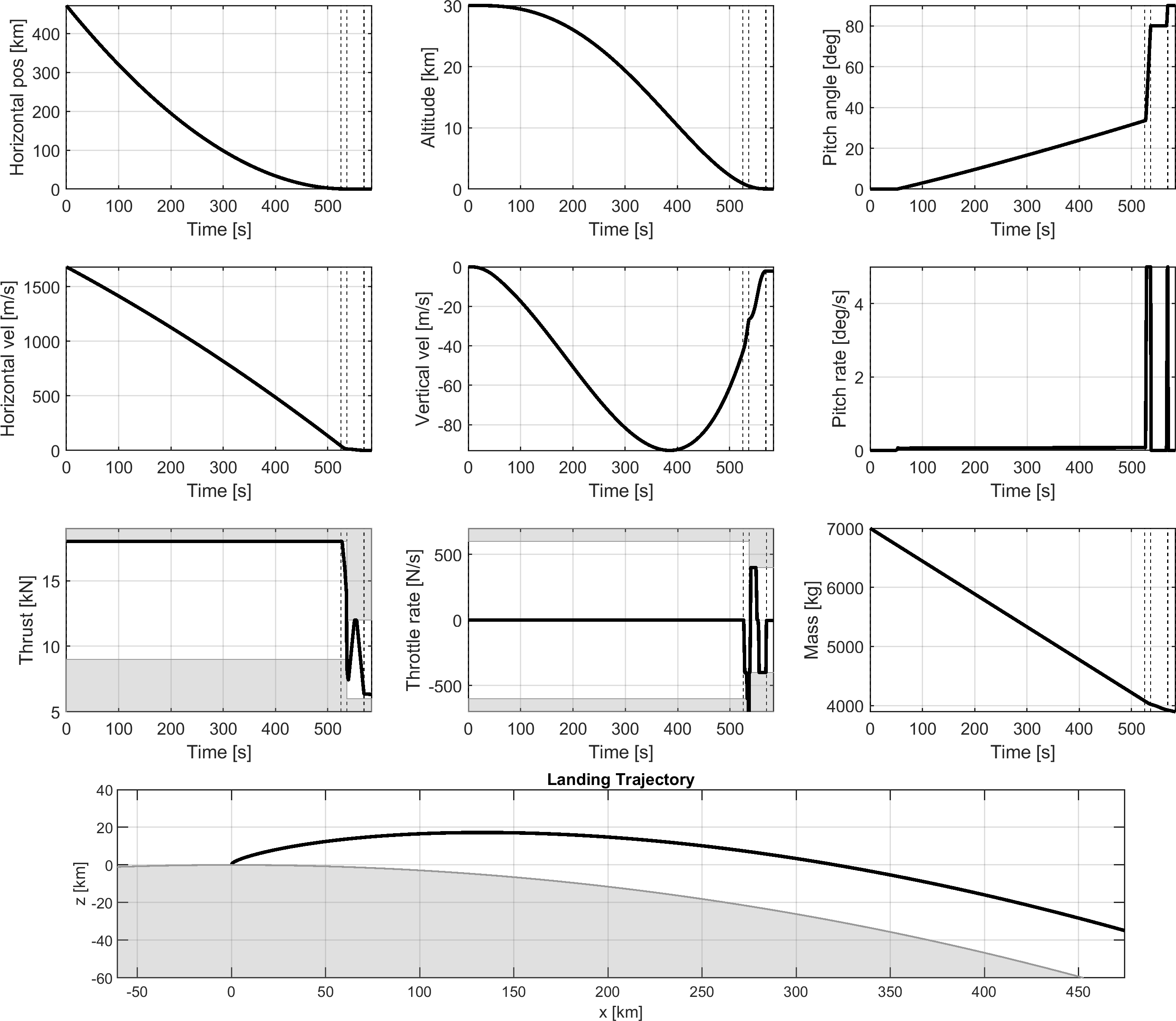}
    \caption{Optimal trajectory -- constant thrust during braking burn. Dotted vertical lines indicate the location of PGA, LGA, and VGA waypoints.}
    \label{fig:optTraj}
\end{figure} 

\begin{table}[!h]
    \centering
    \begin{tabular}{lccccccccc}
      Waypoint  & \coltit{Time\\{[}s{]}} & \coltit{Altitude\\{[}m{]}} & \coltit{Downrange \\ pos. [m]} & \coltit{Vertical \\vel. [m/s]} & \coltit{Horizontal \\vel. [m/s]} & \coltit{Pitch\\ angle [deg]} & \coltit{Mass\\{[}kg{]}}\\\hline
    \textbf{MBB} &  0     & 30000  & 472230.0  & 0     & 1681.6 & 0    & 7000.0\\
    \textbf{PGA} &  525.1 & 920    & 497.1     & -43.2 & 42.1   & 33.5 & 4079.4\\
    \textbf{LGA} &  536.7 & 500    & 214.7     & -26.8 & 13.4   & 80   & 4020.7 \\
    \textbf{VGA} &  569.4 & 30     & 0         & -2    & 0      & 90   & 3923.6\\
    \textbf{MECO} & 584.4 & 0      & 0         & -2    & 0      & 90   & 3894.2\\ \hline
    \end{tabular}
    \caption{Nominal optimal trajectory timeline.}
    \label{tab:optwp}
\end{table}
}

\subsection{Divert manoeuvres}
The computation of optimal divert manoeuvres can also be handled by the same OCP framework used for the nominal trajectory. The only remarkable difference is that divert manoeuvres do not require a multi-phase approach because they are planned after the extinction of the central engine. Therefore, a single OCP phase starting at a given altitude and terminating at VGA is sufficient to model and compute diverts. Full state and command continuity is imposed at the beginning of each divert, to ensure a feasible trajectory. The time of flight is still considered as a free parameter to be optimized. Results obtained for the optimal divert manoeuvres are presented in Section \ref{cmp}.

\section{Candidate on-board guidance solutions}\label{sec3}
The optimal trajectory presented so far serves as reference scenario to confirm feasibility, consolidate the D\&L concept of operations, and identify the best achievable performance in terms of propellant mass consumption. The question that now arises is how to translate it into a feasible, simple, and verifiable on-board guidance law. All the most relevant possibilities that could be used to solve trajectory optimization problems on-board are discussed hereafter. 

\paragraph{On-board trajectory optimization}
This approach would require the on-board implementation and solution of the full-fledged optimal control problem as presented in Section \ref{sec2}, taking into account all the constraints imposed on the trajectory of the vehicle. The resulting trajectories would be fuel optimal and would fulfill all constraints, assuming that a feasible solution (and an appropriate initial guess) exists for the specific (estimated) state of the vehicle when the guidance is executed. To this date, this approach is considered beyond what the current spacecraft state-of-the-art capabilities are, both in terms of computational capabilities and algorithmic ones, the latter being affected by the eternal questions of validation and convergence.

\paragraph{On-board trajectory optimization via sequential convexification}
The full optimization problem can be convexified in a local neighborhood of a given initial solution and solved on-board by a convex solver \cite{cvxx}, whose computational and convergence properties can be bounded, and with a potential limited impact on the optimality and feasibility of the final solution. Nevertheless, the successive convexifications required to handle non-convex constraints (in our case: the nonlinear dynamics of the system, pitch rate constraints, and free final time) would simply displace convergence issues from the solver to the outer convexification loop. If state-of-the-art solutions are converging towards some very promising feasible implementations of this approach  \cite{doi:10.2514/6.2020-0844} -- particularly for fixed final time scenarios where lossless convexification is possible, and notably for rocket landings \cite{spacex} -- this is still considered as a challenging approach in terms of V\&V and industrial implementation. 

\paragraph{Optimal trajectories database}
This is a viable approach where optimal trajectories are computed offline prior to flight, taking into account a set of realistic dispersions on key mission parameters (e.g., initial state, initial mass, engine parameters). The envelope of trajectories is then stored on-board and interpolated on-the-fly, as also suggested in \cite{doi:10.2514/6.2008-6426}, on the basis of the current estimated vehicle's state, and the chosen reference trajectory tracked in closed-loop by the GNC subsystem. Ideally, the bigger the database, the smaller is the fuel penalty with respect to an optimal solution. Unfortunately, the database can very quickly grow in size, especially in the Argonaut case, where up to two divert manoeuvres may be needed. Additionally, this approach automatically results in a quantization of the attainable landing site position, which could only be chosen among the ones included in the database. If this latter issue is less of a concern for Argonaut, the on-board storage of the database may represent a bigger problem, as Argonaut also requires a potentially large terrain features database for Terrain Relative Navigation, whose usefulness is key to meet the challenging landing accuracy requirements of the mission. 
 
\paragraph{Deep reinforcement learning}
Another alternative and quite innovative approach is to implement deep reinforcement learning techniques on-board to compute the optimal trajectory and commands. Currently, this is a quite active research topic, with a few applications to spacecraft guidance already available in the literature \cite{hov,izzo1,izzo2}. But, as for on-board optimization, key issues in explainability and V\&V need to be solved before a possible real-life implementation on a space mission.

\paragraph{Motion planning}
Motion planning approaches are usually adopted in robotics to solve very constrained trajectory optimization problems, potentially with multiple agents evolving in a very cluttered and highly dynamic environment. Examples of sampling-based techniques applied to space mission scenarios can be found in the literature \cite{Surovik_Scheeres_2015,CAPOLUPO20178279}, but there is still a significant leap to be made for a future concrete on-board application, specifically in the available vs. required computational capabilities for these methods to provide a sufficiently good, feasible solution.

\paragraph{Semi-analytic sub-optimal approach}
The semi-analytic approach exploits some simple guidance solutions that have been already flown on past missions, such as the powered explicit guidance, the polynomial guidance, or the Apollo guidance. These solutions are unable to explicitly account for constraints and generally do not provide a fuel-optimal trajectory. To make them work on a real scenario, the trajectory's waypoints are specifically engineered to ensure feasibility and mimic the optimal solution despite the use of simpler algorithms. Waypoints' location and states are optimized offline to fulfill mission and system constraint and to reduce as much as possible the inevitable fuel penalty that this approach entails. 
A similar solution has recently flown on Chang'E missions \cite{changee,change}, and represents a possible viable option also for Argonaut. The objectives of the next sections are to derive a dedicated semi-analytic guidance solution and sub-optimal trajectory design process for Argonaut, and to analyze whether or not the resulting propellant mass penalty would preclude its future implementation.

\section{Guidance algorithms \& sub-optimal trajectory design}\label{sec4}
The design of a sub-optimal trajectory starts with the definition of appropriate trajectory waypoints, and with the identification of suitable guidance algorithms that can solve the two-point boundary value problem between waypoints. Here it is proposed to reuse the same waypoints already presented in the mission scenario description, i.e., MBB, PGA, LGA, VGA and MECO. The different guidance solutions retained for braking burn, pitch-up, and powered descent \& diverts are presented in the following sections. The vertical descent and landing phase being a trivial constant speed rectilinear vertical motion, its description is omitted from the discussion. Finally, we outline a process to optimize the degrees of freedom of the trajectory, taking explicitly into account the chosen end-to-end guidance approach from MBB to touchdown.

\subsection{Braking burn guidance}\label{bbg}
The guidance solution proposed for the braking burn phase is strongly inspired by the Powered Explicit Guidance (PEG) algorithm \cite{peg1} flown on the Space Shuttle \cite{Mchenry1979SpaceSA}, and commonly implemented to compute the fuel-optimal ascent trajectory of a launch vehicle. There are many variants of PEG that work with different simplifying hypotheses and models, some of which have also been successfully implemented onboard lunar landers \cite{change}. In the proposed approach we consider a given constant thrust $T$ along the direction ${\bm{u}} = \left[u_r,u_{\phi},u_{\theta}\right]^\top$ with constant $I_s$ and mass flow rate. We use the same Keplerian dynamics in spherical coordinates  used for the optimal trajectory, but with a small change in the expression of the thrust vector to account for its newly introduced formulation. Velocity derivatives of Eq.~\ref{polardyn} are modified as follows
\begin{subequations}
\begin{align}
\dot v_r & = \frac{v_{\phi}^2 + v_{\theta}^2}{r} - \frac{\mu}{r^2} + \frac{T}{m}u_r\\
\dot v_{\phi} & = -\frac{v_{\phi}v_r}{r} - \frac{v^2_{\theta}}{r}\tan\phi  + \frac{T}{m}u_{\phi}\\
\dot v_{\theta} & = -\frac{v_{\theta}v_r}{r} + \frac{v_{\phi}v_{\theta}}{r}\tan\phi + \frac{T}{m}u_{\theta}
\end{align}
\end{subequations}
where the mass of the vehicle varies linearly with time 
\begin{equation}m\left(t\right) = m_0 - \frac{T}{g_0I_s} \left(t - t_0\right)\end{equation}
The objective of the braking burn algorithm is to provide the appropriate steering law ${\bm{u}}\left(t\right)$ that allows reaching a desired final state at $t_f$ from a given initial state at $t_0$. To the best of our knowledge, there is no closed-form solution to this problem. Nevertheless, when considering a uniform gravity field, the Pontryagin's Maximum Principle states \cite{lawden1963optimal,pingg} that the optimal thrust direction is given by the well known \emph{bilinear tangent law} (see also Appendix \ref{pegflat})
\begin{equation}{\bm{u}} = \frac{{\bm{c}}t + {\bm{b}}}{\left\|{\bm{c}}t + {\bm{b}}\right\|}\end{equation}
where ${\bm{b}} = \left[b_r,b_{\phi},b_{\theta}\right]^\top$ and ${\bm{c}} = \left[c_r,c_{\phi},c_{\theta}\right]^\top$ are two appropriate constant vectors. In our case a bilinear law does not guarantee an optimal fuel consumption, as our gravity field is not uniform, but Keplerian. The choice of a non-uniform gravity field model is dictated by Argonaut mission profile: the initial altitude and orbital velocity of the vehicle are such that the trajectory spans many hundreds of kilometers along-track, with a time of flight of several minutes. The use of a uniform gravity field for the computation of a descent trajectory would severely affect the accuracy of the solution, and, consequently, would have a significant impact on the amount of propellant spent by the GNC to keep the vehicle on such a trajectory. Fortunately, as it can also be observed on the optimized trajectory profile, the bilinear tangent law can be still considered as a very good approximation of the optimal steering law, and, therefore, it is selected as baseline for the braking burn guidance.

To compute the two above-mentioned constants, let's first address the design of the trajectory, and let the on-board implementation considerations for later. The boundary conditions of our manoeuvring problem can be mathematically stated as follows
\begin{subequations}
\begin{align}
    \Delta t = t_f - t_0 & = \text{free}\\
    r\left(t_0\right) & = r_0\\
    \phi\left(t_0\right) & = \phi_0\\
    \theta \left(t_0\right) & =  \text{free}\\
    {\bm{v}}\left(t_0\right)  & = {\bm{v}}_0\\
    {\bm{x}}\left(t_f\right) & = {\bm{x}}_f
\end{align}
\end{subequations}
where ${\bm{x}}:=\left[r,\phi,\theta,v_r,v_{\phi},v_{\theta}\right]^\top$, and ${\bm{v}} := \left[v_r,v_{\phi},v_{\theta}\right]^\top$. This means that we require the vehicle to reach a desired final state starting from the periselene of a given orbit, but without imposing the argument of periselene of the orbit (defined with respect to the landing site frame introduced in Fig.~\ref{fig:ldFrames}) nor the time of flight -- exactly as done to derive the optimal trajectory. Since $\theta\left(t_0\right)$ is a free parameter, we can set $c_{\theta} = 0$, as commonly done in PEG for a uniform gravity field with free initial (or final) downrange position (see Appendix \ref{pegflat}). We can also recognize that multiplying both constants by the a positive scalar value does not affect the steering law. Therefore, knowing that the initial direction of the thrust vector is very likely to have a negative component along the tangential axis, we can set $b_{\theta} = -1$. The guidance problem becomes now a simple nonlinear root-finding problem where five parameters ${\bm{y}} := \left[\Delta t,c_{\phi},c_r,b_{\phi},b_r\right]^\top$ are adjusted to match five desired initial conditions ${\bm{z}}_0 := \left[r_0,\phi_0,v_{r_0},v_{\phi_0},v_{\theta_0}\right]^\top$. The error at $t_0$ is obtained by numerically propagating the trajectory back in time starting from the desired final state ${\bm{x}}_f$. The back-propagated target state vector $\tilde {\bm{z}}(t_0;{\bm{y}},{\bm{x}}_f)$ is compared to the desired target states ${\bm{z}}_0$, and the parameters adjusted iteratively as follows
\begin{equation}{\bm{y}}^{\left(k\right)} = {\bm{y}}^{\left(k-1\right)} - J^{-1}\left[\tilde {\bm{z}}(t_0;{\bm{y}}^{\left(k-1\right)},{\bm{x}}_f) - {\bm{z}}_0 \right]\end{equation}
with $J = \partial \tilde {\bm{z}}/\partial {\bm{y}}$ being the Jacobian of the target state vector with respect to the parameters. A purely tangent steering law with $c_r = c_{\phi} = b_r = b_{\phi} = 0$ can be used as initial guess to ensure a quick convergence. The initial guess for the time of flight $\Delta t$ can be computed via the Tsiolkovsky equation knowing the given thrust level $T$ and $I_s$, the initial mass of the vehicle, and using as velocity increment $\Delta v = \left\| {\bm{v}}_f - {\bm{v}}_0\right\|$.
In the tests executed in the frame of this work a few (typically less than 5) iterations are sufficient to converge towards a feasible solution. 
Note that the thrust level $T$ is not considered as an adjustment parameter, but it is set to its maximum allowable value to minimize the propellant consumption, as the optimal trajectory suggests.

\paragraph{Degrees of freedom for trajectory design} The degrees of freedom to be used for trajectory design are the following states at PGA: $r$, $\theta$, $v_{r}$, and $v_{\theta}$, representing a subset of the final desired states ${\bm{x}}_f$ that the braking burn algorithm is targeting. All the remaining parameters are specified as boundary conditions (as $r_0$, $\phi_0$, $v_{\theta_0}$, $v_{\phi_0}$, $\phi_f$, and $v_{\phi_f}$), directly specified in the algorithm (as the thrust value $T = T_{\max}$), or an outcome of the algorithm itself (as the time of flight $\Delta t$ and initial along-track angular position $\theta_0$).

\begin{figure}
    \centering
    \includegraphics[scale=0.7]{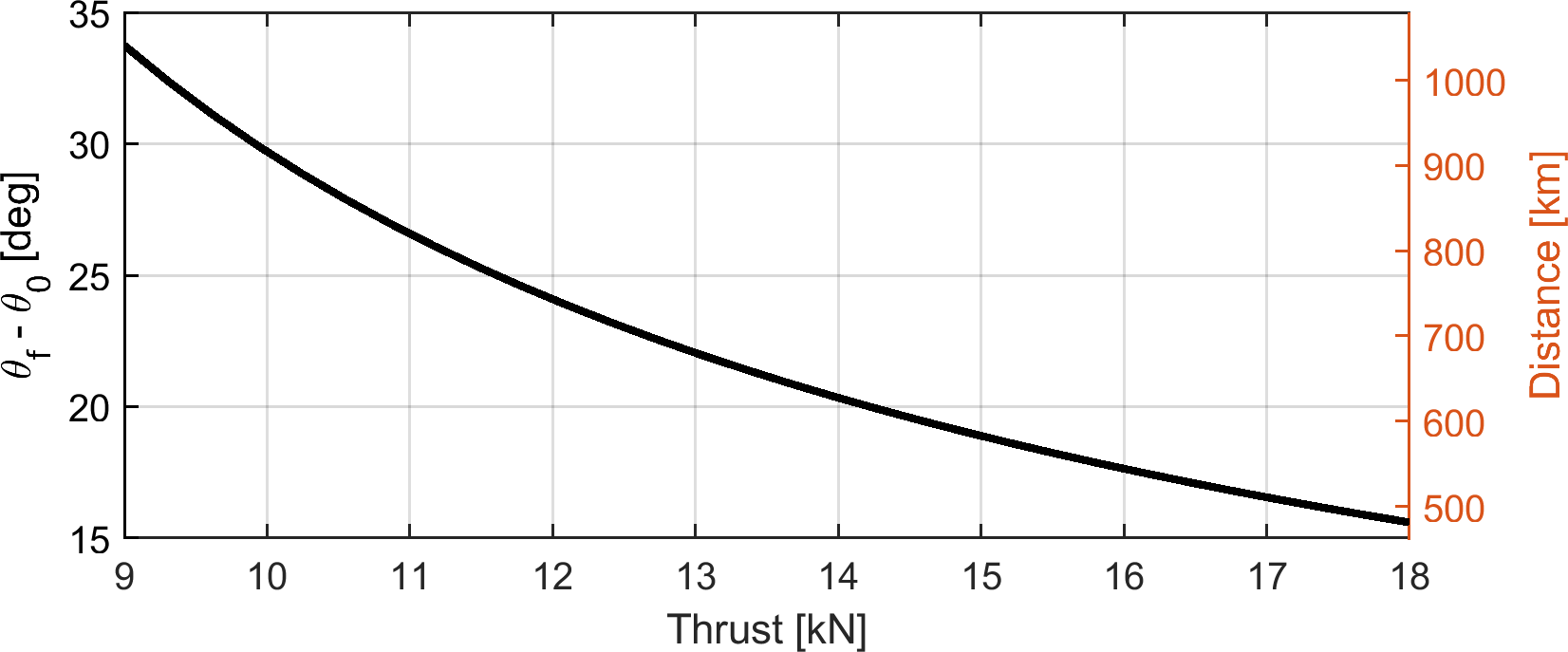}
    \caption{Reachable $\theta\left(t_0\right)$ as function of the thrust level $T$ for $\theta_f = 90$ deg. Notice that 1 degree corresponds to $\approx$30 km of along-track displacement at 30 km of altitude.}
    \label{fig:thrustTheta}
\end{figure}

\paragraph{Enforcing $\theta\left(t_0\right)$ for on-board implementation} When implementing the braking burn guidance algorithm on-board, we cannot anymore let $\theta\left(t_0\right)$ be a free parameter. On a realistic scenario, the vehicle begins its descent from a given $\theta\left(t_0\right)$, potentially far from its nominal value determined in the design phase. The algorithm must correct for this initial error and still bring the vehicle to the desired final state ${\bm{x}}_f$.
If we are allowed to choose the constant thrust level $T$ between a minimum and maximum bound $T_{\min}$ and $T_{\max}$, we can also enforce a desired $\theta\left(t_0\right)$. We can easily compute offline the reachable $\theta\left(t_0\right)$ as a function of the thrust level $T$ (as in the example shown in Fig.~\ref{fig:thrustTheta}), and, once on-board, interpolate the precomputed values based on the current estimated vehicle state. Alternatively, we can modify the root-finding problem of the braking burn algorithm adding $T$ to the list of parameters ${\bm{y}}$ and requiring that all six final states ${\bm{z}}_f \equiv {\bm{x}}_f$ are matched. In this case we can simply propagate forward in time the initial state and match the end state to the desired one: the initial state being completely set, there is no further need to back-propagate the trajectory from the final state.

\subsection{Pitch-up guidance}
The pitch-up guidance consists in a predefined attitude manoeuvre that reorients the main engine thrust direction towards the local vertical. Given an initial (estimated) direction ${\bm{u}}_0$ at $t_0$ and a desired final direction ${\bm{u}}_f$ at $t_f$, we rotate the thrust direction around the following axis 
\begin{equation}{\bm{p}} = \frac{{\bm{u}}_0\times {\bm{u}}_f}{\left\|{\bm{u}}_0\times {\bm{u}}_f\right\|}\end{equation}
for an angle 
\begin{equation}\Delta\psi\left(t\right) = - \dot \psi_{\max} \left(t - t_0\right)\end{equation}
so that the steering command can be written as
\begin{equation}{\bm{u}}\left(t\right) = \exp\left\{\Delta\psi\left(t\right)\crossmat{{\bm{p}}}\right\} {\bm{u}}_0\end{equation}
The duration of the manoeuvre is dictated by the maximum steering rate $\dot \psi_{\max}$ of the vehicle
\begin{equation}\Delta t =t_f- t_0 = \frac{1}{\dot \psi_{\max}}\cos^{-1} \left({{\bm{u}}^\top_f {\bm{u}}_0 }\right)\end{equation}
At the same time we reduce the thrust level as follows
\begin{equation}T\left(t\right) = T\left(t_0\right) - \dot T_{\max}\left(t - t_0\right)\end{equation}
to prepare for the powered descent and divert manoeuvres. In the proposed implementation, the final desired pitch angle is set to 80 degrees, as required at LGA. The numerical integration of the equations of motion with the thrust vector profile provided by the pitch-up guidance constitutes the desired trajectory to be followed during this phase.

\paragraph{Degrees of freedom for trajectory design} there are no degrees of freedom to be used for trajectory design in this phase, as all the parameters are either taken as initial conditions from the braking burn phase, as for the states at PGA, or directly enforced by design in the guidance algorithm, as the thrust magnitude and steering profile. The final state at the end of the pitch-up phase is entirely obtained from (and constrained by) the numerical integration of the pitch-up guidance profile with the given initial conditions.

\subsection{Powered descent and diverts}
To realize the powered descent phase and its two diverts we rely on a polynomial guidance algorithm \cite{acik1}, similar to what has flown on the Apollo missions \cite{apollo}. This time we use Cartesian coordinates to describe the translational dynamics of the lander
\begin{equation}\ddot {\bm{r}} = \frac{T}{m}{\bm{u}} + {\bm{g}}\left({\bm{r}}\right) =: {\bm{a}}\end{equation}
where the vector ${\bm{a}}$, equal to the sum of gravity and specific thrust, is referred to as total acceleration. We choose a cubic function of time to enforce initial states, end states, as well as initial and final total accelerations
\begin{equation}{\bm{a}}\left(t\right) = {\bm{a}}_0 + {\bm{C}}_1 t + {\bm{C}}_2 t^2 + {\bm{C}}_3 t^3\end{equation}
The enforcement of end conditions in terms of acceleration, velocity and position results in a linear system of equations
\begin{subequations}
\begin{align}
{\bm{a}}_f & = {\bm{a}}_0 + {\bm{C}}_1 t_f + {\bm{C}}_2 t_f^2 + {\bm{C}}_3 t_f^3 \\
{\bm{v}}_f & = {\bm{v}}_0 + {\bm{a}}_0 t_f + \frac{1}{2}{\bm{C}}_1 t_f^2 + \frac{1}{3}{\bm{C}}_2 t_f^3 + \frac{1}{4}{\bm{C}}_3 t_f^4\\
{\bm{r}}_f & = {\bm{r}}_0 + {\bm{v}}_0 t_f + \frac{1}{2}{\bm{a}}_0 t_f^2 + \frac{1}{6}{\bm{C}}_1 t_f^3 + \frac{1}{12}{\bm{C}}_2 t_f^4 + \frac{1}{20}{\bm{C}}_3 t_f^5
\end{align}
\end{subequations}
whose analytical solution is
\begin{equation}
\mymatrix{{\bm{C}}_1\\{\bm{C}}_2\\{\bm{C}}_3} = \mymatrix{\frac{3}{t_f}I & -\frac{24}{t_f^2}I &   \frac{60}{t_f^3}I\\
 -\frac{12}{t_f^2}I&  \frac{84}{t_f^3}I& -\frac{180}{t_f^4}I\\
 \frac{10}{t_f^3}I & -\frac{60}{t_f^4}I&  \frac{120}{t_f^5}I} \mymatrix{{\bm{a}}_f - {\bm{a}}_0\\
{\bm{v}}_f - {\bm{v}}_0 - {\bm{a}}_0 t_f \\
{\bm{r}}_f - {\bm{r}}_0 - {\bm{v}}_0 t_f - \frac{1}{2}{\bm{a}}_0 t_f^2}
\end{equation}
The main engine thrust vector is then computed by subtracting the gravitational acceleration from the total acceleration vector.
The enforcement of total acceleration vectors at the beginning and at the end of the phase is very important to ensure thrust continuity between pitch-up and powered descent, and to prepare the vehicle for the vertical descent phase where ${\bm{a}}_f = 0$. 

\begin{figure}
    \centering
    \includegraphics[scale = 0.6]{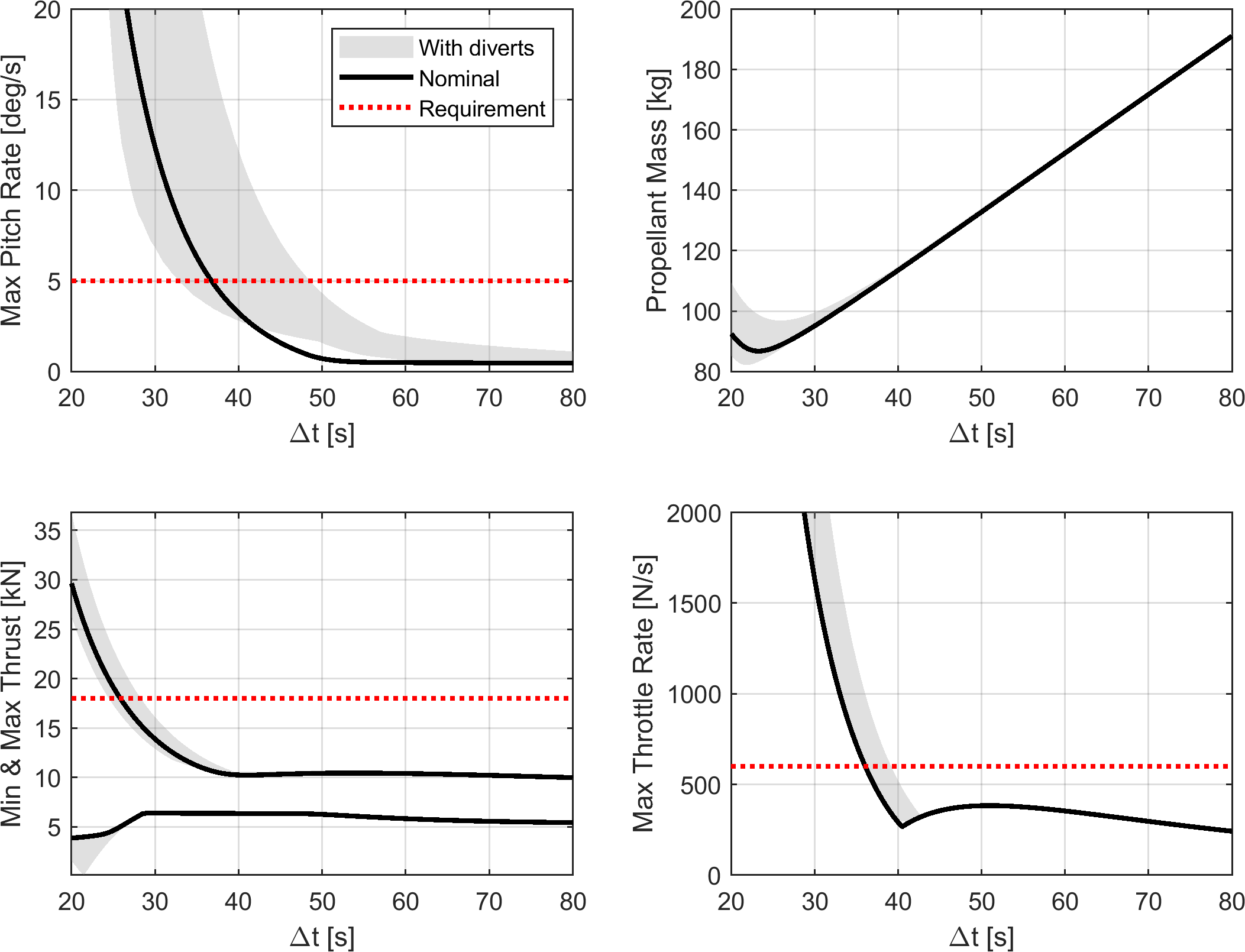}
    \caption{Sensitivity of powered descent and diverts parameters to the time of flight $\Delta t$}
    \label{fig:sensPwr}
\end{figure}

The overall duration of the manoeuvre $\Delta t = t_f - t_0$ can be considered as a fixed parameter to be optimized off-line when designing the trajectory. Its choice obviously influences the overall $\Delta v$ of the trajectory, but, more importantly, the smoothness of the resulting thrust vector, thus the fulfillment of thrust and steering rate constraints. Fig.~\ref{fig:sensPwr} shows an example of how these key powered descent parameters vary with the chosen time of flight, highlighting that, for the nominal trajectory (solid black line) of the given example, throttle rate constraint are the most stringent ones.

The very same algorithm is used for divert manoeuvres, with the initial state of a divert manoeuvre being the estimated vehicle state at the instant of divert. The time of flight of a divert manoeuvre $\Delta t_{div}$ can be seen an additional off-line optimization parameter function of the amplitude of the divert. The same analysis of how the time of flight influences the trajectory parameters can be repeated for diverts, in order to determine a suitable time of flight for diverts manoeuvres. Fig.~\ref{fig:sensPwr} shows the effects of diverts (grey shaded areas) for the test-cases considered in Section \ref{cmp}, and supposing that the time of flight between LGA and VGA is not modified in case of divert. It can be seen that, as expected, diverts have a significant impact on the maximum pitch rate of the vehicle, and that in some divert cases pitch rate constraints can only be fulfilled by increasing the allowed time to reach VGA. 

\paragraph{Degrees of freedom for trajectory design} The two degrees of freedom of this phase that can be exploited in the trajectory design are the times of flight with and without diverts, $\Delta t_{div}$ and $\Delta t$.

\subsection{Trajectory parameters optimization}
Summarizing the phase-by-phase analysis presented above, the design parameters that completely define the nominal sub-optimal trajectory are the $\left(r,\theta,v_r,v_{\theta}\right)$ states at PGA and the time of flight $\Delta t$ of the powered descent phase. The choice of their values is a key off-line process that makes the use of the proposed end-to-end on-board guidance strategy possible. An in-house implementation of a Differential Evolution (DE) algorithm \cite{DEpaper}, able to solve nonlinear, non-smooth, constrained parametric  optimization problems, is used with the goal of minimizing the overall propellant mass consumption and ensuring the fulfillment of trajectory constraints. 
To help the optimizer converge, we include the optimal reference trajectory parameters in the initial random population of the optimizer, even if no initial guess is strictly needed for DE to provide a solution. Steering rate, throttle rate, thrust bounds, and pitch, altitude, and velocity constraints at LGA are included in the definition of the fitness function of the problem, together with the propellant mass objective, as to penalize infeasible solutions. Once the nominal (i.e., without diverts) trajectory is found, the time of flight of each divert manoeuvre of our reference test cases (as defined in the next section) is also optimized with the same process, following the concepts highlighted by the sensitivity analysis of Figure \ref{fig:sensPwr}.

\begin{table}[!h]
    \centering
    \begin{tabular}{lccccccccc}
      Waypoint  & \coltit{Time\\{[}s{]}} & \coltit{Altitude\\{[}m{]}} & \coltit{Downrange \\ pos. [m]} & \coltit{Vertical \\vel. [m/s]} & \coltit{Horizontal \\vel. [m/s]} & \coltit{Pitch\\ angle [deg]} & \coltit{Mass\\{[}kg{]}}\\\hline
    \textbf{MBB} &  0     & 30000  & 473906.4  & 0     & 1681.6 & -4.7 & 7000.0\\
    \textbf{PGA} &  526.7 & 898.3  & 622.9     & -38.5 & 38.7   & 31.9 & 4070.7\\
    \textbf{LGA} &  539.7 & 498.2  & 312.4     & -24.3 & 17.2   & 80   & 4016.9 \\
    \textbf{VGA} &  579.3 & 30     & 0         & -2    & 0      & 90   & 3908.8\\
    \textbf{MECO} & 594.3 & 0      & 0         & -2    & 0      & 90   & 3879.5\\ \hline
    \end{tabular}
    \caption{Nominal sub-optimal trajectory timeline.}
    \label{tab:suboptwp}
\end{table}

\begin{figure}[!htb]
    \centering
    \includegraphics[width=\textwidth]{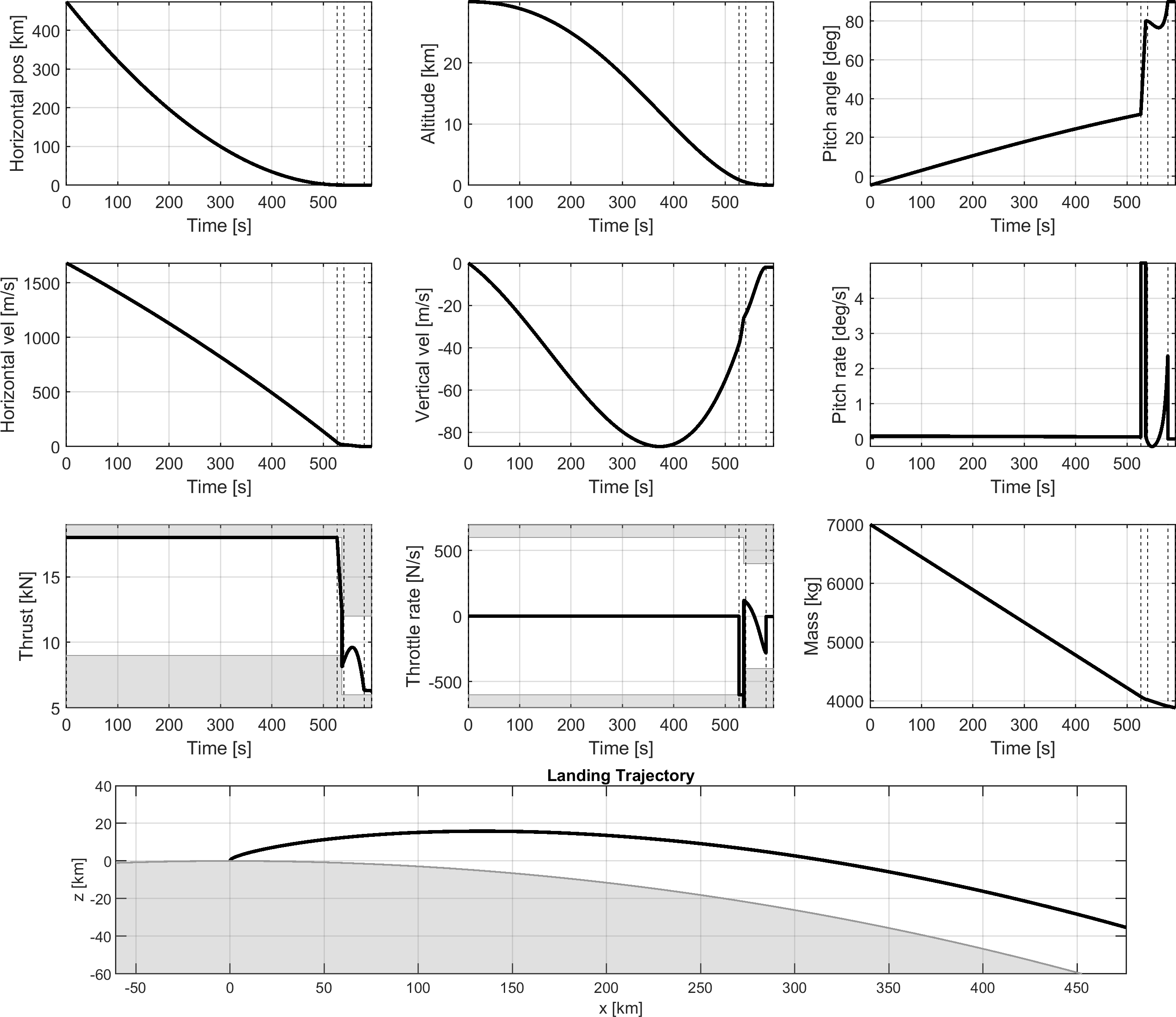}
    \caption{Sub-optimal trajectory. Dotted vertical lines indicate the location of PGA, LGA, and VGA waypoints.}
    \label{fig:suboptTraj}
\end{figure}

\begin{figure}[t]
    \centering
    \begin{subfigure}[b]{0.47\textwidth}
        \includegraphics[width=\textwidth]{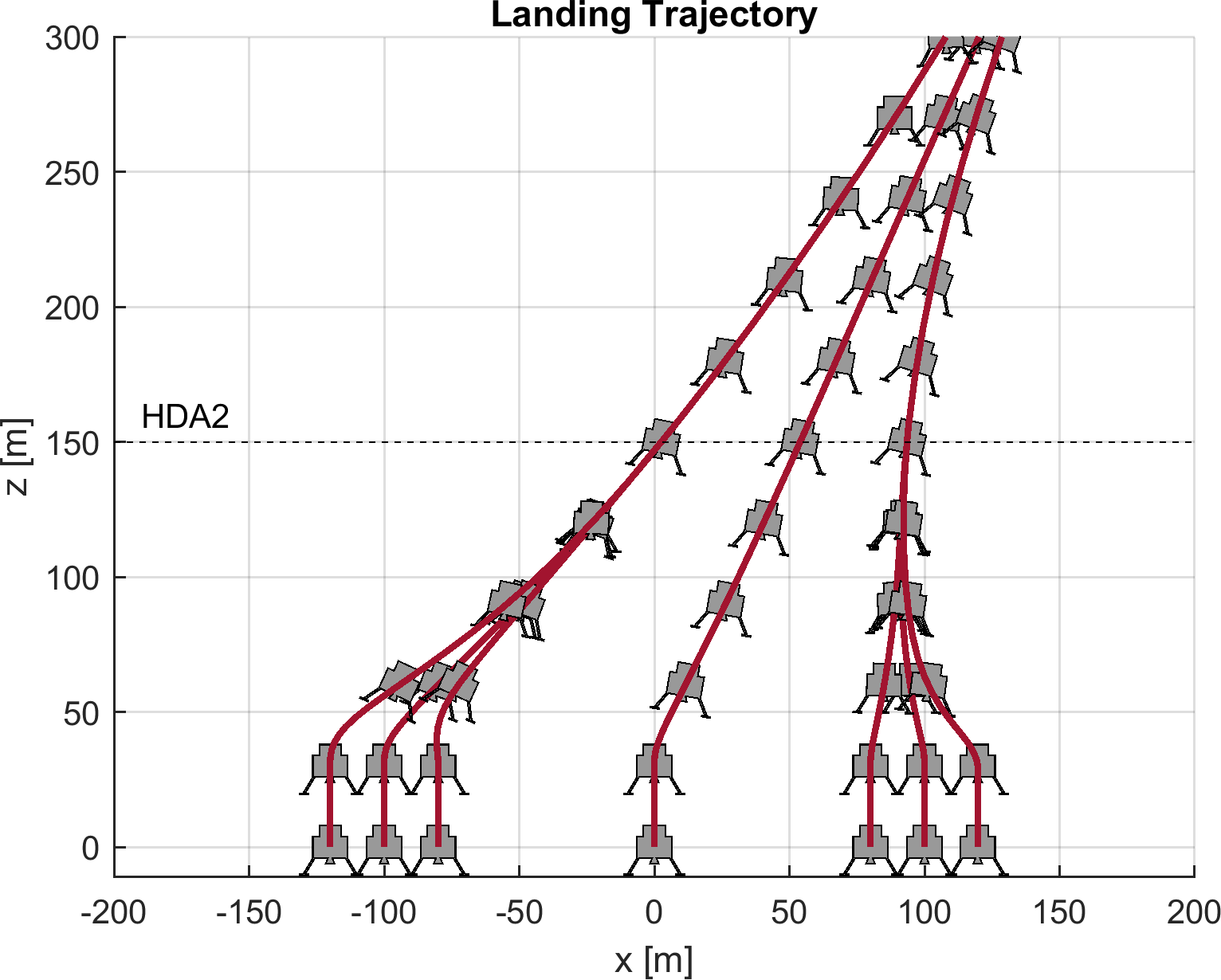}
        \caption{Optimal trajectory}
        \label{fig:optttTraj}
    \end{subfigure}
    \hfill
    \begin{subfigure}[b]{0.47\textwidth}
        \includegraphics[width=\textwidth]{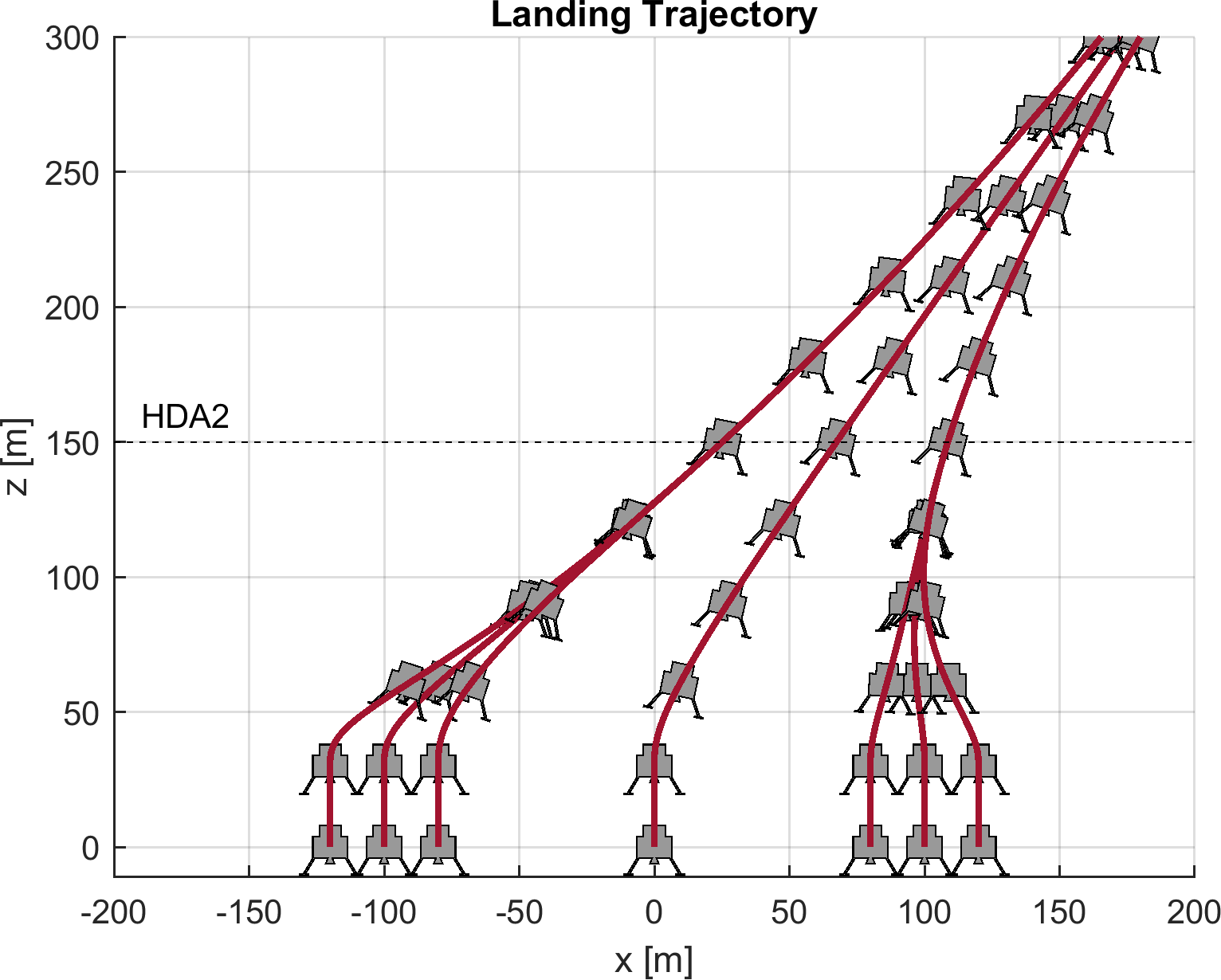}
        \caption{Sub-optimal trajectory}
        \label{fig:analyticTraj}
    \end{subfigure}
    \caption{Final divert and vertical D\&L trajectories.}
    \label{fig:comptraj}
\end{figure}

Table \ref{tab:suboptwp} provides the final results for the nominal sub-optimal trajectory, while Figure \ref{fig:suboptTraj} depicts the evolution of states and thrust profile over time as well as the final trajectory in the landing site-centered inertial reference frame. All parameters are within their allowed range, confirming that the proposed guidance and optimization approach is able to generate a feasible trajectory. Waypoints states are close, but not exactly equal to the optimal ones. Figure \ref{fig:comptraj}, comparing the final portion of landing of sub-optimal and optimal trajectories, shows that the sub-optimal trajectory has a slightly shallower approach to ground, but a smoother pitch profile when reaching VGA. Overall, the fuel consumption and time of flight are only slightly increased with respect to optimal, with a propellant mass penalty of 14.7 kg for a 9.9 s longer trajectory. 

It is worth underlining that an alternative possibility, not pursued in the frame of this work, would have been to let DE optimize the propellant consumption of the worst-case trajectory with divert. This could result in a higher nominal mass consumption (which would no longer be the optimization criteria to be used) but would potentially guarantee a lower worst-case mass consumption among all diverted trajectories.

\subsection{On-board implementation}
The braking burn and powered descent algorithms allow for a closed-loop implementation where trajectories and commands are recomputed at each GNC cycle, thus providing closed-loop position control capabilities. 
Nevertheless, when the vehicle approaches the endpoint of a phase (PGA for braking burn, and VGA for powered descent), both algorithms become numerically unstable as the remaining time of flight approaches zero. Some workarounds exist to overcome this issue \cite{wmjpl}, one of which is to design the next trajectory phase to be executed before reaching the end of the previous one.
An alternative possibility for an on-board implementation consists in executing the two algorithms only once at the beginning of the corresponding phases, and then track the computed (open-loop) trajectory with a closed-loop position and attitude control system. In this case, the thrust commands computed by the guidance module serve as feed-forward signal to the feedback controller to improve the trajectory tracking performance.

\section{Comparative analysis}\label{cmp}
The two trajectories, optimal and sub-optimal, are compared on seven different test cases. Each test case includes a different divert manoeuvre at HDA1 and HDA2, as shown in Table \ref{tabbb}. The size of each divert is set equal to the maximum allowed one, i.e., $\pm$100 m for HDA1 and $\pm$20 m for HDA2. Only along-track diverts are considered in this work, giving rise to 6 possible combinations of forward (F) and backward (B) diverts, plus one nominal trajectory (N). The propellant mass consumption of each phase and of the whole trajectory is reported in Table \ref{tabbb}. Please note that the location and velocity of LGA in the optimal trajectory is slightly different than the one in the sub-optimal trajectory, and therefore, the phase-by-phase comparison can only be done qualitatively. 

Results show that the additional propellant mass required by the sub-optimal strategy is well below 1\% of the optimal propellant mass consumption for all selected test-cases. The backward (B) case is found to be the worst-case scenario, with 20.2 extra kilograms of propellant needed on top of what the fuel optimal trajectory requires. We can also notice that, qualitatively, the braking burn algorithm consumption is quite close to the optimal one, which confirms our assumption on the suitability of a bilinear tangent law for the braking burn phase. The braking burn propellant representing more than 95\% of the total propellant spent during D\&L, this is a very important outcome in view of a future consolidation of the GNC design. 

\begin{figure}[t]
    \centering
    \begin{subfigure}[b]{\textwidth}
        \includegraphics[width=\textwidth]{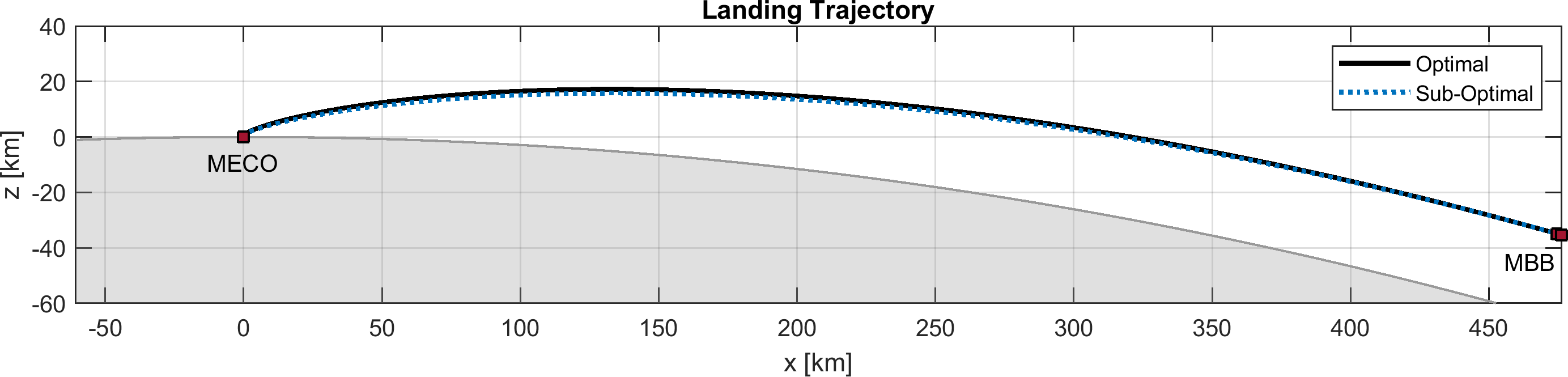}
        \caption{MBB to MECO}
        \label{fig:tc2}
    \end{subfigure}
    \\\vspace{0.3cm}
    \begin{subfigure}[b]{0.7\textwidth}
        \includegraphics[width=\textwidth]{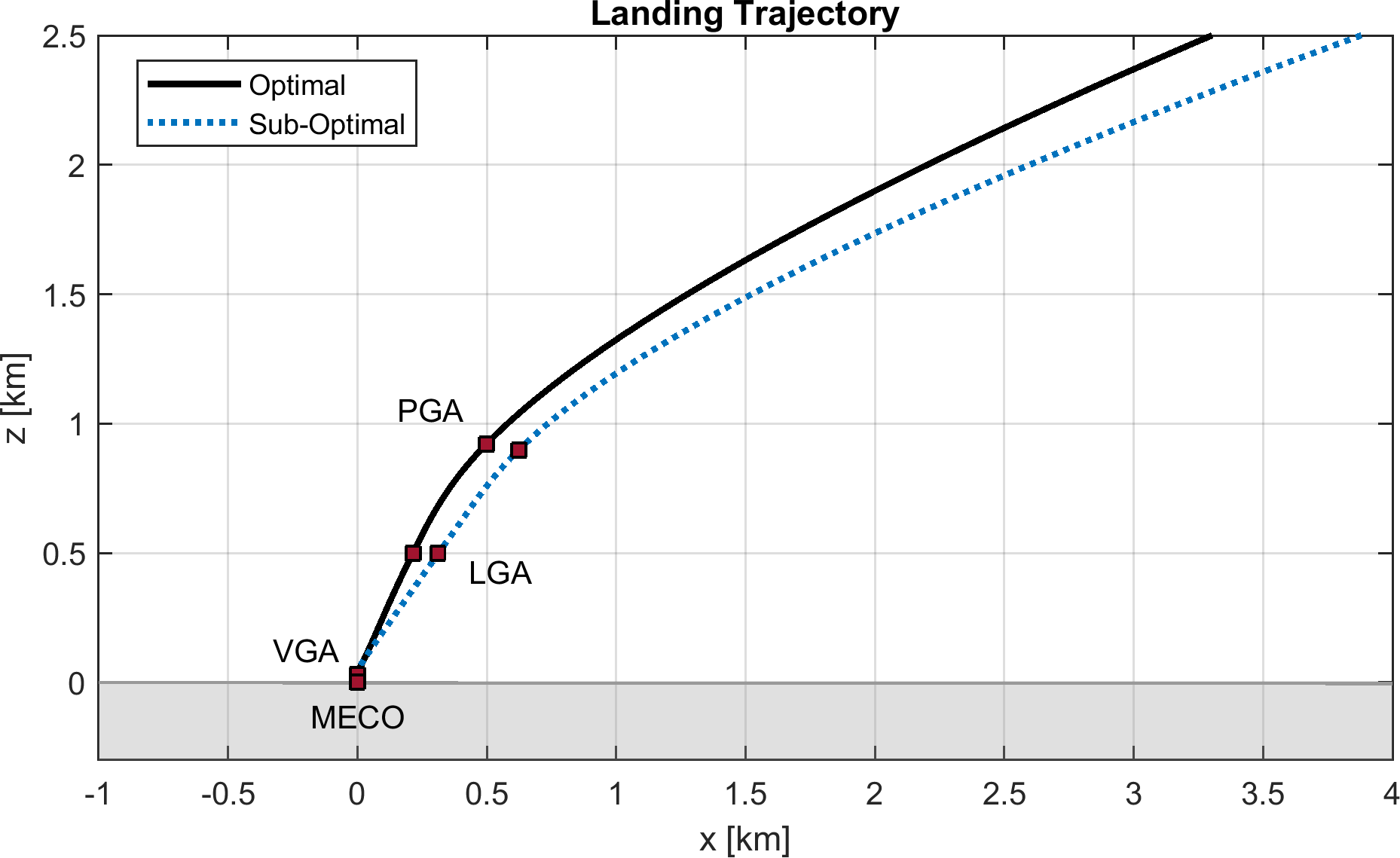}
        \caption{PGA to MECO}
        \label{fig:tc1}
    \end{subfigure}
    \caption{Optimal vs. sub-optimal nominal trajectories.}
    \label{fig:comptraj0}
\end{figure}

Polynomial guidance results are also quite close to optimal, with no excessive increase in the propellant expenditure when performing diverts. There are two reasons that can explain this result. Firstly, the parameters of the sub-optimal trajectory are explicitly optimized to minimize the fuel expenditure, despite the use of a non-optimal polynomial guidance. Secondly, the diverts required by Argonaut have a quite limited amplitude if compared to what is commonly found in the literature \cite{acik1} for similar applications, and, therefore, they still allow for the use of a non-optimal guidance law with marginal impact on the mass budget.

While a more detailed analysis would be required in phase B2 to further consolidate the propellant figures presented in this work and extend the test-cases to the whole possible range of diverts, these preliminary results already highlight the potential of the proposed approach. The use of simple guidance algorithms, combined with the off-line, end-to-end, and "algorithm-aware" optimization of trajectory parameters, allows for a very limited mass penalty despite a very simple on-board algorithmic implementation. 

\section{Conclusions}\label{secc}
This study has outlined a suitable optimal landing and descent approach for the Argonaut mission. The fuel-optimal trajectory has been computed accounting for mission and vehicle constraints, confirming mission feasibility and establishing a preliminary sizing of the mass that can be landed on the Moon's surface. A simple end-to-end guidance approach for the whole descent \& landing trajectory, including divert manoeuvres, has been also presented. The main contributions of this work are the adaptation of existing powered explicit guidance algorithms to Argonaut, as well as the definition of an off-line trajectory optimization process that minimizes propellant budget and fulfills mission and system constraints despite the use of sub-optimal guidance laws.
The comparison between fuel-optimal and sub-optimal trajectories on several test-cases, with and without diverts, has shown that the mass penalty of the sub-optimal approach is less than 1\% of the fuel-optimal propellant mass. The conclusion is that the proposed guidance design has no severe effect on the payload mass capabilities of Argonaut, and, therefore, can be considered as a promising candidate solution for implementation in phase B2/C/D, potentially yielding to minimal development risk and V\&V effort when compared to on-board optimization-based solutions, and to minimal (to none) on-board storage and memory footprint when compared to an on-board trajectory database.

\begin{landscape}
\begin{table}[]
\begin{center}
\begin{tabular}{llcccccccccc}
\multicolumn{2}{c}{\textbf{Test-Case}} & \begin{tabular}[c]{@{}c@{}}{Divert}\\ HDA1 {[}m{]}\end{tabular}  & \begin{tabular}[c]{@{}c@{}}{Divert}\\ HDA2 {[}m{]}\end{tabular}  & \begin{tabular}[c]{@{}c@{}}{Time of}\\ flight {[}s{]}\end{tabular} & \begin{tabular}[c]{@{}c@{}}Braking Burn\\ \& Pitch-Up {[}kg{]}\end{tabular} & \begin{tabular}[c]{@{}c@{}}Powered\\ Descent {[}kg{]}\end{tabular} & \begin{tabular}[c]{@{}c@{}}Vertical\\ Descent {[}kg{]}\end{tabular} & \textbf{\begin{tabular}[c]{@{}c@{}}Total\\ {[}kg{]}\end{tabular}} & \begin{tabular}[c]{@{}c@{}} Equivalent \\ $\Delta V$ {[}m/s{]}\end{tabular} & \begin{tabular}[c]{@{}c@{}} Prop. mass \\ penalty {[}kg{]}\end{tabular}  \\\hline\hline
\multirow{7}{*}{\STAB{\rotatebox[origin=c]{90}{\textit{OPTIMAL}}}}
 & \textbf{N}  & 0      & 0     & 584.4 & 2979.3 & 97.1 & 29.4 & \textbf{3105.8} & 1898.4 & --\\
 & \textbf{F}  & $-100$ & 0     & 585.2 & 2979.3 & 99.9 & 29.4 & \textbf{3108.7} & 1900.8 & --\\
 & \textbf{FF} & $-100$ & $-20$ & 585.8 & 2979.3 & 101.4 & 29.4 & \textbf{3110.1}& 1902.0 & --\\
 & \textbf{FB} & $-100$ & $+20$ & 585.0 & 2979.3 & 99.1 & 29.4 & \textbf{3107.8} & 1900.1 & --\\
 & \textbf{B}  & $+100$ & 0     & 584.1 & 2979.3 & 97.8 & 29.4 & \textbf{3106.6} & 1899.1 & --\\
 & \textbf{BF} & $+100$ & $-20$ & 584.1 & 2979.3 & 97.7 & 29.4 & \textbf{3106.4} & 1898.9 & --\\
 & \textbf{BB} & $+100$ & $+20$ & 584.4 & 2979.3 & 98.8 & 29.4 & \textbf{3107.5} & 1899.8 & --\\ \hline
 \multirow{7}{*}{\STAB{\rotatebox[origin=c]{90}{\textit{SUB-OPTIMAL}}}}
& \textbf{N} & 0 & 0 & 594.3 & 2974.2 & 117.1 & 29.3 & \textbf{3120.5} &  1910.1 & 14.7 \hspace{0.2cm}(+0.47\%)\\
& \textbf{F} & $-100$ & 0 & 597.1 & 2974.2 & 122.8 & 29.3 & \textbf{3126.2} &  1914.9 & 17.5 \hspace{0.2cm}(+0.56\%)\\
& \textbf{FF} & $-100$ & $-20$ & 598.6 & 2974.2 & 125.6 & 29.3 & \textbf{3129.1} &  1917.2 & 19.0 \hspace{0.2cm}(+0.61\%)\\
& \textbf{FB} & $-100$ & $+20$ & 595.6 & 2974.2 & 120.1 & 29.3 & \textbf{3123.5} &  1912.6 & 15.7 \hspace{0.2cm}(+0.51\%)\\
& \textbf{B} & $+100$ & 0 & 597.1 & 2974.2 & 123.4 & 29.3 & \textbf{3126.8} &  1915.4 & 20.2 \hspace{0.2cm}(+0.65\%)\\
& \textbf{BF} & $+100$ & $-20$ & 595.6 & 2974.2 & 120.2 & 29.3 & \textbf{3123.7} &  1912.7 & 17.3 \hspace{0.2cm}(+0.56\%)\\
& \textbf{BB} & $+100$ & $+20$ & 595.6 & 2974.2 & 121.0 & 29.3 & \textbf{3124.5} &  1913.4 & 17.0 \hspace{0.2cm}(+0.55\%)\\\hline
\end{tabular}
\end{center}
\caption{Propellant consumption.}
\label{tabbb}
\end{table}
\end{landscape}

\appendix
\section{Powered Explicit Guidance for a flat planet}\label{pegflat}
We approximate the dynamics of the lander by considering a point mass flying over a flat planet with a uniform gravity field along $z$. We also consider a constant thrust $T$ in a given direction ${\bm{u}}$ with constant $I_{sp}$ and mass flow rate $\dot m_p$, so that the equations of motion are
\[\ddot {\bm{r}} = {\bm{g}} + \gamma{\bm{u}}\]
with $\gamma := T/m$. As we seek to minimize the overall fuel consumption and knowing that our propulsion system has a constant propellant flow rate, we choose as cost function the manoeuvre duration, $J = t_f$. Starting from a given initial state at $t_0 = 0$, we impose a final desired velocity, final desired lateral and vertical positions, and we let the time of flight and final downrange position be free parameters
\begin{align*}
    {\bm{x}}\left(t_0\right) & = {\bm{x}}_0\\
    t_f & = \text{free}\\
    x\left(t_f\right) & = \text{free}\\
    y\left(t_f\right) & = y_f\\
    z\left(t_f\right) & = z_f\\
    \dot {\bm{r}}\left(t_f\right) & = {\bm{v}}_f
\end{align*}
To compute the optimal trajectory, we exploit the Pontryagin's Maximum Principle. The Hamiltonian of this constrained optimization problem is written as
\[\mathcal{H} = {\bm{\psi}}_r^\top \dot {\bm{r}} + {\bm{\psi}}_v^\top \left({\bm{g}} + \gamma {\bm{u}}\right)\]
which, knowing that $\gamma > 0$, is maximized when ${\bm{u}}$ is aligned with ${\bm{\psi}}_v$ (also known as \emph{primer vector}) and $\gamma$ is at its maximum value. Since ${\bm{u}}$ is, by definition, a unitary vector, we set
\[{\bm{u}} = \frac{{\bm{\psi}}_v}{\left\|{\bm{\psi}}_v\right\|}\]
The dynamics of the co-state are derived from the Hamiltonian, yielding to
\[\begin{split}
\dot {\bm{\psi}}_r & = 0\\
\dot {\bm{\psi}}_v & = -{\bm{\psi}}_r
\end{split}\]
This first order ODE system can be integrated analytically by introducing two constant vectors ${\bm{b}}$ and ${\bm{c}}$
\[\begin{split}
{\bm{\psi}}_r & = -{\bm{c}}\\
{\bm{\psi}}_v & = {\bm{c}}t + {\bm{b}}
\end{split}\]
The optimal thrust direction is then given by the well known \emph{bilinear tangent law}
\[{\bm{u}}\left(t\right) = \frac{{\bm{c}}t + {\bm{b}}}{\left\|{\bm{c}}t + {\bm{b}}\right\|}\]
The Hamiltonian can be rewritten as
\[\mathcal{H} = -{\bm{c}}^\top \dot {\bm{r}} + \left({\bm{c}}t + {\bm{b}}\right)^\top {\bm{g}} + \gamma\left(t\right) \left\|{\bm{c}}t + {\bm{b}}\right\|\]
Transversality conditions on the Hamiltonian and on the co-states yield to the two following useful expressions
\[\begin{split}
\mathcal{H}\left(t_f\right) & = 1 = -{\bm{c}}^\top {\bm{v}}_f + \left({\bm{c}}t_f + {\bm{b}}\right)^\top {\bm{g}} + \gamma\left(t_f\right) \left\|{\bm{c}}t_f + {\bm{b}}\right\|\\
\psi_{r_x}\left(t_f\right) & = 0 = -c_x
\end{split}\]
The last equation tells us that we can remove $c_x$ from the set of parameters to be found to construct the optimal trajectory and set it directly to $c_x = 0$. If we integrate the trajectory in time, and we add the condition on $\mathcal{H}\left(t_f\right)$, we end up with a set of 6 equations
\begin{align}
y_0 - y_f + v_{0} t_f + \frac{1}{2}g_y t_f^2 +  \int\limits_{0}^{t_f} \int\limits_{0}^t \gamma\left(\tau\right) u_y\left(\tau,{\bm{b}},c_y,c_z\right) d\tau dt & = 0\\
z_0 - z_f + w_{0} t_f + \frac{1}{2}g_z t_f^2 +  \int\limits_{0}^{t_f} \int\limits_{0}^t \gamma\left(\tau\right) u_z\left(\tau,{\bm{b}},c_y,c_z\right) d\tau dt & = 0\\
{\bm{v}}_0 - {\bm{v}}_f + {\bm{g}} t_f + \int\limits_0^{t_f} \gamma\left(t\right) {\bm{u}}\left(t,{\bm{b}},c_y,c_z\right) dt &  = 0 \\
-{\bm{c}}^\top {\bm{v}}_f + \left({\bm{c}}t_f + {\bm{b}}\right)^\top {\bm{g}} + \gamma\left(t_f\right) \left\|{\bm{c}}t_f + {\bm{b}}\right\| - 1 & = 0
\end{align}
with 6 unknowns: $t_f$, ${\bm{b}}$, $c_y$ and $c_z$. Notice that only the $y$ and $z$ position coordinates have been considered, since $x\left(t_f\right) = \text{free}$. The problem of finding the optimal command is therefore translated into a nonlinear multivariate root finding problem ${\bf{f}}\left({\bm{y}}\right) = 0$ in the parameters ${\bm{y}} = \left(t_f,{\bm{b}},c_y,c_z\right)$.  

\nocite{*}
\bibliography{biblio}

\begin{thebibliography}{34}
\newcommand{\enquote}[1]{``#1''}
\providecommand{\natexlab}[1]{#1}
\providecommand{\url}[1]{\texttt{#1}}
\providecommand{\urlprefix}{URL }
\expandafter\ifx\csname urlstyle\endcsname\relax
  \providecommand{\doi}[1]{\discretionary{}{}{}https://doi.org/#1}\else
  \providecommand{\doi}[1]{\discretionary{}{}{}\urlstyle{rm}\url{https://doi.org/#1}}\fi

\bibitem[{Nilsson et~al.(2022)Nilsson, Rometsch, Casini, Guerra, Becker,
  Treuer, de~Medeiros, Schnellbaecher, Vock, and Cowley}]{3dmodel}
Nilsson, T., Rometsch, F., Casini, A. E.~M., Guerra, E., Becker, L., Treuer,
  A., de~Medeiros, P., Schnellbaecher, H., Vock, A., and Cowley, A.,
  \enquote{Using Virtual Reality to Design and Evaluate a Lunar Lander: The EL3
  Case Study,} \emph{Extended Abstracts of the 2022 CHI Conference on Human
  Factors in Computing Systems}, Association for Computing Machinery, New York,
  NY, USA, 2022.
\newblock \doi{10.1145/3491101.3519775}.

\bibitem[{Klumpp(1971)}]{apollo}
Klumpp, A.~R., \enquote{Apollo Lunar-Descent Guidance,} Tech. Rep. {R-695}, MIT
  Charles Stark Draper Laboratory, 1971.

\bibitem[{Betts(1998)}]{betts}
Betts, J.~T., \enquote{Survey of Numerical Methods for Trajectory
  Optimization,} \emph{Journal of Guidance, Control, and Dynamics}, Vol.~21,
  No.~2, 1998, pp. 193--207.
\newblock \doi{10.2514/2.4231}.

\bibitem[{W{\"a}chter and Biegler(2006)}]{ipopt}
W{\"a}chter, A., and Biegler, L.~T., \enquote{On the implementation of an
  interior-point filter line-search algorithm for large-scale nonlinear
  programming,} \emph{Mathematical Programming}, Vol. 106, No.~1, 2006, pp.
  25--57.
\newblock \doi{10.1007/s10107-004-0559-y}.

\bibitem[{Malyuta et~al.(2022)Malyuta, Reynolds, Szmuk, Lew, Bonalli, Pavone,
  and Açıkmeşe}]{cvxx}
Malyuta, D., Reynolds, T.~P., Szmuk, M., Lew, T., Bonalli, R., Pavone, M., and
  Açıkmeşe, B., \enquote{Convex Optimization for Trajectory Generation: A
  Tutorial on Generating Dynamically Feasible Trajectories Reliably and
  Efficiently,} \emph{IEEE Control Systems Magazine}, Vol.~42, No.~5, 2022, pp.
  40--113.
\newblock \doi{10.1109/MCS.2022.3187542}.

\bibitem[{Reynolds et~al.(2020)Reynolds, Malyuta, Mesbahi, A\c{c}\i{}kme\c{s}e,
  and Carson}]{doi:10.2514/6.2020-0844}
Reynolds, T., Malyuta, D., Mesbahi, M., A\c{c}\i{}kme\c{s}e, B., and Carson,
  J.~M., \emph{A Real-Time Algorithm for Non-Convex Powered Descent Guidance},
  2020.
\newblock \doi{10.2514/6.2020-0844}.

\bibitem[{Blackmore(2016)}]{spacex}
Blackmore, L., \enquote{Autonomous Precision Landing of Space Rockets,}
  \emph{The Bridge, National Academy of Engineering}, Vol.~46, No.~4, 2016.

\bibitem[{A\c{c}\i{}kme\c{s}e et~al.(2008)A\c{c}\i{}kme\c{s}e, Scharf,
  Blackmore, and Wolf}]{doi:10.2514/6.2008-6426}
A\c{c}\i{}kme\c{s}e, B., Scharf, D., Blackmore, L., and Wolf, A.,
  \enquote{{Enhancements on the Convex Programming Based Powered Descent
  Guidance Algorithm for Mars Landing},} \emph{AIAA/AAS Astrodynamics
  Specialist Conference and Exhibit}, 2008.
\newblock \doi{10.2514/6.2008-6426}.

\bibitem[{Hovell and Ulrich(2020)}]{hov}
Hovell, K., and Ulrich, S., \emph{On Deep Reinforcement Learning for Spacecraft
  Guidance}, 2020.
\newblock \doi{10.2514/6.2020-1600}.

\bibitem[{S{\'a}nchez and Izzo(2016)}]{izzo1}
S{\'a}nchez, C.~M., and Izzo, D., \enquote{Optimal real-time landing using deep
  networks,} 2016.

\bibitem[{Willis et~al.(2016)Willis, Izzo, and Hennes}]{izzo2}
Willis, S., Izzo, D., and Hennes, D., \enquote{Reinforcement Learning for
  Spacecraft Maneuvering Near Small Bodies,} 2016.

\bibitem[{Surovik and Scheeres(2015)}]{Surovik_Scheeres_2015}
Surovik, D., and Scheeres, D., \enquote{Heuristic Search and Receding-Horizon
  Planning in Complex Spacecraft Orbit Domains,} \emph{Proceedings of the
  International Conference on Automated Planning and Scheduling}, Vol.~25,
  No.~1, 2015, pp. 291--295.
\newblock \doi{10.1609/icaps.v25i1.13694},
  \urlprefix\url{https://ojs.aaai.org/index.php/ICAPS/article/view/13694}.

\bibitem[{Capolupo et~al.(2017)Capolupo, Simeon, and Berges}]{CAPOLUPO20178279}
Capolupo, F., Simeon, T., and Berges, J.-C., \enquote{Heuristic Guidance
  Techniques for the Exploration of Small Celestial Bodies,}
  \emph{IFAC-PapersOnLine}, Vol.~50, No.~1, 2017, pp. 8279--8284.
\newblock \doi{https://doi.org/10.1016/j.ifacol.2017.08.1401}, 20th IFAC World
  Congress.

\bibitem[{Zhang et~al.(2021)Zhang, Li, Wang, and Guan}]{changee}
Zhang, H., Li, J., Wang, Z., and Guan, Y., \enquote{{Guidance Navigation and
  Control for Chang'E-5 Powered Descent},} \emph{Space: Science \& Technology},
  2021.
\newblock \doi{10.34133/2021/9823609}.

\bibitem[{Li et~al.(2016)Li, Jiang, and Tao}]{change}
Li, S., Jiang, X., and Tao, T., \enquote{{Guidance Summary and Assessment of
  the Chang'E-3 Powered Descent and Landing},} \emph{Journal of Spacecraft and
  Rockets}, Vol.~53, No.~2, 2016, pp. 258--277.
\newblock \doi{10.2514/1.A33208}.

\bibitem[{Jaggers(1977)}]{peg1}
Jaggers, R., \enquote{An explicit solution to the exoatmospheric powered flight
  guidance and trajectory optimization problem for rocket propelled vehicles,}
  \emph{Guidance and Control Conference}, 1977.
\newblock \doi{10.2514/6.1977-1051}.

\bibitem[{McHenry et~al.(1979)McHenry, Long, Cockrell, Thibodeau, and
  Brand}]{Mchenry1979SpaceSA}
McHenry, R.~L., Long, A., Cockrell, B.~F., Thibodeau, J.~R., and Brand, T.~J.,
  \enquote{{Space Shuttle Ascent Guidance, Navigation, and Control},} \emph{The
  Journal of the Astronautical Sciences}, Vol. XXVII, No.~1, 1979, pp. 1--38.

\bibitem[{Lawden(1963)}]{lawden1963optimal}
Lawden, D.~F., \emph{Optimal Trajectories for Space Navigation}, Butterworths,
  London, 1963.

\bibitem[{Lu and Callan(2023)}]{pingg}
Lu, P., and Callan, R., \enquote{Propellant-Optimal Powered Descent Guidance
  Revisited,} \emph{Journal of Guidance, Control, and Dynamics}, Vol.~46,
  No.~2, 2023, pp. 215--230.
\newblock \doi{10.2514/1.G007214}.

\bibitem[{Ploen et~al.(2006)Ploen, A\c{c}\i{}kme\c{s}e, and Wolf}]{acik1}
Ploen, S., A\c{c}\i{}kme\c{s}e, B., and Wolf, A., \enquote{{A Comparison of
  Powered Descent Guidance Laws for Mars Pinpoint Landing},} \emph{AIAA/AAS
  Astrodynamics Specialist Conference and Exhibit}, 2006.
\newblock \doi{10.2514/6.2006-6676}.

\bibitem[{Storn and Price(1997)}]{DEpaper}
Storn, R., and Price, K., \enquote{Differential Evolution -- A Simple and
  Efficient Heuristic for global Optimization over Continuous Spaces,}
  \emph{Journal of Global Optimization}, Vol.~11, No.~4, 1997, pp. 341--359.
\newblock \doi{10.1023/A:1008202821328}.

\bibitem[{Wong et~al.(2002)Wong, Masciarelli, and Singh}]{wmjpl}
Wong, E., Masciarelli, J., and Singh, G., \enquote{{Autonomous Guidance and
  Control Design for Hazard Avoidance and Safe Landing on Mars},} \emph{AIAA
  Atmospheric Flight Mechanics Conference and Exhibit}, 2002.
\newblock \doi{10.2514/6.2002-4619}.

\bibitem[{Rea and Bishop(2010)}]{doi:10.2514/6.2010-8026}
Rea, J., and Bishop, R., \enquote{{Analytical Dimensional Reduction of a Fuel
  Optimal Powered Descent Subproblem},} \emph{AIAA Guidance, Navigation, and
  Control Conference}, 2010.
\newblock \doi{10.2514/6.2010-8026}.

\bibitem[{Delporte and Sauvinet(1992)}]{doi:10.2514/6.1992-1145}
Delporte, M., and Sauvinet, F., \enquote{{Explicit guidance law for manned
  spacecraft},} \emph{Aerospace Design Conference}, 1992.
\newblock \doi{10.2514/6.1992-1145}.

\bibitem[{Lu et~al.(2012)Lu, Forbes, and Baldwin}]{doi:10.2514/6.2012-4843}
Lu, P., Forbes, S., and Baldwin, M., \enquote{{A Versatile Powered Guidance
  Algorithm},} \emph{AIAA Guidance, Navigation, and Control Conference}, 2012.
\newblock \doi{10.2514/6.2012-4843}.

\bibitem[{Sostaric and Rea(2005)}]{doi:10.2514/6.2005-6287}
Sostaric, R., and Rea, J., \enquote{{Powered Descent Guidance Methods For The
  Moon and Mars},} \emph{AIAA Guidance, Navigation, and Control Conference and
  Exhibit}, 2005.
\newblock \doi{10.2514/6.2005-6287}.

\bibitem[{Lee et~al.(2010)Lee, Ely, Sostaric, Strahan, Riedel, Ingham,
  Wincentsen, and Sarani}]{doi:10.2514/6.2010-7717}
Lee, A., Ely, T., Sostaric, R., Strahan, A., Riedel, J., Ingham, M.,
  Wincentsen, J., and Sarani, S., \enquote{{Preliminary Design of the Guidance,
  Navigation, and Control System of The Altair Lunar Lander},} \emph{AIAA
  Guidance, Navigation, and Control Conference}, 2010.
\newblock \doi{10.2514/6.2010-7717}.

\bibitem[{Lu(2018)}]{doi:10.2514/1.G003243}
Lu, P., \enquote{{Propellant-Optimal Powered Descent Guidance},} \emph{Journal
  of Guidance, Control, and Dynamics}, Vol.~41, No.~4, 2018, pp. 813--826.
\newblock \doi{10.2514/1.G003243}.

\bibitem[{A\c{c}\i{}kme\c{s}e and Ploen(2007)}]{doi:10.2514/1.27553}
A\c{c}\i{}kme\c{s}e, B., and Ploen, S.~R., \enquote{{Convex Programming
  Approach to Powered Descent Guidance for Mars Landing},} \emph{Journal of
  Guidance, Control, and Dynamics}, Vol.~30, No.~5, 2007, pp. 1353--1366.
\newblock \doi{10.2514/1.27553}.

\bibitem[{D'Souza(1997)}]{doi:10.2514/6.1997-3709}
D'Souza, C., \enquote{An optimal guidance law for planetary landing,}
  \emph{Guidance, Navigation, and Control Conference}, 1997.
\newblock \doi{10.2514/6.1997-3709}.

\bibitem[{Topcu et~al.(2005)Topcu, Casoliva, and
  Mease}]{doi:10.2514/6.2005-6286}
Topcu, U., Casoliva, J., and Mease, K., \emph{Fuel Efficient Powered Descent
  Guidance for Mars Landing}, 2005.
\newblock \doi{10.2514/6.2005-6286}.

\bibitem[{Kamath et~al.(2022)Kamath, Elango, Yu, Mceowen, Carson, and
  Açıkmeşe}]{kamath2022realtime}
Kamath, A.~G., Elango, P., Yu, Y., Mceowen, S., Carson, J.~M., and Açıkmeşe,
  B., \enquote{Real-Time Sequential Conic Optimization for Multi-Phase Rocket
  Landing Guidance,} , 2022.

\bibitem[{Szmuk et~al.(2020)Szmuk, Reynolds, and
  A\c{c}\i{}kme\c{s}e}]{doi:10.2514/1.G004549}
Szmuk, M., Reynolds, T.~P., and A\c{c}\i{}kme\c{s}e, B., \enquote{Successive
  Convexification for Real-Time Six-Degree-of-Freedom Powered Descent Guidance
  with State-Triggered Constraints,} \emph{Journal of Guidance, Control, and
  Dynamics}, Vol.~43, No.~8, 2020, pp. 1399--1413.
\newblock \doi{10.2514/1.G004549}.

\bibitem[{Scharf et~al.(2015)Scharf, Ploen, and
  Acikmese}]{doi:10.2514/6.2015-0850}
Scharf, D.~P., Ploen, S.~R., and Acikmese, B.~A., \emph{Interpolation-Enhanced
  Powered Descent Guidance for Onboard Nominal, Off-Nominal, and Multi-X
  Scenarios}, 2015.
\newblock \doi{10.2514/6.2015-0850}.

\end{thebibliography}
\end{document}